\documentclass{nature}
\usepackage{caption}
\usepackage{graphicx}
\usepackage{subcaption}
\makeatletter
\let\saved@includegraphics\includegraphics
\AtBeginDocument{\let\includegraphics\saved@includegraphics}
\renewenvironment*{figure}{\@float{figure}}{\end@float}
\usepackage[export]{adjustbox}
\usepackage{cleveref}
\title{Characterizing regimes of hydrodynamic escape of close-in low mass exoplanets}
\author {J. H. Guo$^{1,2,3}$}
\begin{document}

\maketitle

\begin{affiliations}
 \item Yunnan Observatories,
Chinese Academy of Sciences, P.O. Box 110, Kunming 650011, China
 \item University of CAS, Beijing, China
 \item Key Laboratory for the Structure and Evolution of
Celestial Objects, CAS, Kunming 650011, China.
\end{affiliations}

\begin{abstract}

The hydrodynamic escape driven by external or internal energy sources sculpts the population of low mass close-in planets. However, distinguishing between the driving mechanisms responsible for the hydrodynamic escape of hydrogen-rich atmospheres is a complex task due to the involvement of many physical factors. My simulations show that the hydrodynamic escape can be driven solely by thermal energy deposited in the lower layers of the atmosphere due to the heat flux originating from the planetary core or bolometric heating from the star even in the absence of other energy sources, as long as the planet's Jeans parameter is below 3. Otherwise, stellar extreme ultraviolet irradiation or tidal forces are necessary in driving the escape, which means that the Jeans parameter is incapable of distinguishing the driving mechanisms, as it is only related to the properties of planet. Here, an upgraded Jeans parameter that takes into account tidal forces is introduced, which allows us to accurately categorize the driving mechanisms. The results show that when the upgraded Jeans parameter falls below 3 or exceeds 6, the atmospheric escape is primarily driven by tidal forces or extreme ultraviolet radiation from the host star, respectively. In the range of 3 to 6, both factors can trigger the escape of the atmosphere. I find that planets with high gravitational potential and low stellar irradiation are more likely to undergo subsonic escape, although transonic escape is prevalent among most planets. Moreover, the ionization status is significantly dependent on the gravitational potential. The upgraded Jeans parameter, which is closely related to the underlying physics, provides a concise method to categorize the driving mechanisms of hydrodynamic escape. The results can be applied to planetary evolution calculations.
\end{abstract}

The escape of hydrogen was first discovered in the atmosphere of HD 209458b$^{1}$. A hydrogen-rich atmosphere can be heated to several thousand Kelvin due to the heating by stellar X-ray and extreme-ultraviolet radiation (XUV)$^{2}$. As a result, the hydrostatic conditions will no longer be valid, causing the atmosphere to dynamically expand and escape into space$^{3-6}$. Such escape can be approximated by the energy-limited equation $^{7-8}$. On the other hand, Kepler mission discovered thousands of planets outside the Solar system with sizes distributed over a wide range$^{9-10}$. In fact, even a small amount of hydrogen and helium can cause a significant increase in the size of planets$^{11}$ owing to the low density. As a result of large radii (or low densities), the atmospheres of planets can be evaporated through hydrodynamic escape$^{12}$ so that their masses and radii are a function of time$^{13-14}$. Fulton et al. $^{10}$ discovered the existence of a valley in the radius distribution of short-period planets smaller than Neptune by the precise measurements of stellar radii and confirmed a bimodal distribution around a radius of $\sim$ 1.5R$_{\oplus}$. Moreover, the study of the planetary population indicates that the valley can be attributed to the loss of the hydrogen-helium envelope due to the intense XUV radiation emitted by the host star$^{13,15-16}$ or cooling luminosity of the core $^{17-18}$. In addition, the valley can also be related to the processes of formation of gas-poor$^{19}$ or water-rich planets$^{20-21}$.

Theoretical models predict three mechanisms for hydrodynamic atmospheric escape: thermal energy (TE)-$^{17,22}$, tidally-$^{23}$, and XUV-driven escape$^{3-6}$. The thermal energy of the atmosphere, provided by the bolometric radiation of its host star or the cooling of the planet, can lead to boil-off$^{22}$ or spontaneous$^{17-18}$ hydrodynamic escape that is independent of other energy sources (here, I refer to the boil-off and spontaneous escape as TE-driven escape). The TE-driven escape can easily occur on planets with high equilibrium temperatures and low gravitational potential, particularly during planetary formation$^{22}$ or when core-luminosity is released$^{18}$. TE-driven escape rates are limited by the sound speed and density at the planetary or Bondi radius (or sonic point)$^{17-18,22}$, although the energy available in the atmosphere can be much greater than the mechanical energy of the escaping wind. Tidally-driven escape$^{23}$ refers to cases where the work done by the stellar tidal forces is more prominent than the effects of other energy sources. For these first two cases, an organized outflow can occur in the atmosphere even if the heating from XUV energy is not considered. For an XUV-driven escape, on the other hand, the energy responsible for driving the escape originates from the stellar XUV irradiation. Moreover, the mass loss rates predicted by the XUV-driven escape are proportional to the XUV fluxes received at the planet's orbit $^{4-6,24}$.

The escape of atmosphere is controlled by various factors, including the planet's mass and radius, the temperature of the atmosphere, the XUV irradiation, and the tidal forces exerted by the host star. In earlier research, the planetary mass, the altitude and temperature at exobase, compose the Jeans parameter $^{25}$ that is used to characterize the manner of escape of the atmosphere for the planets in our solar system$^{26-27}$. However, determining the location of the exobase requires employing numerical modeling, which diminishes the applicability of the Jeans parameter for classification. In this paper, I thus defined the Jeans parameter using the planet's mass, radius, and equilibrium temperature (which is determined by the bolometric luminosity of the host star). This generalized Jeans parameter can be expressed as
\begin{equation}
\lambda=\frac{GM_{\rm p}\mu}{k_{\rm B}TR_{\rm p}},
\end{equation}
where M$_{\rm p}$ is the mass of the planet, R$_{\rm p}$ is the radius of planet, T is equilibrium temperature, G is the gravitational constant, $\mu$ is the mean molecular mass, and $k_{\rm B}$ is the Boltzmann constant. In cases where the heat provided only by stellar optical and infrared radiation (bolometric luminosity) is deposited in the lower layers of the atmosphere, a transition from TE-driven hydrodynamic escape to Jeans escape can occur when the Jeans parameter, predicted by Equation (1), is larger than $\sim$ 2-3.6$^{28}$. The phenomenon is distinct from that of close-in exoplanets heated by XUV radiation, where XUV-driven hydrodynamic escape occurs even in regimes with a Jeans parameter much greater than 3.6$^{5-6;29-31}$.

To date, the atmospheric escape has been observed in many Hot Jupiters and Neptunes by space and ground-based telescopes$^{1,32-34}$ and the observation signals of Jupiter-like planets can be explained by XUV-driven escape$^{35-38}$. Moreover, studies on several Jovian or sub-Jovian planets have suggested a prospective transition from XUV-driven to tidally-driven escape when the ratio of the substellar radius to the polar radius exceeds 1.2 or 1.02$^{23}$. However, investigations of the driving mechanism for the escape of the atmosphere have been inconclusive. The parameter distribution of low-mass planets, commonly attributed to atmospheric evaporation, can be explained by both XUV- and TE-driven escape$^{39}$. Identifying the escape mechanism is important because the mass loss predicted by different driving mechanisms can vary in orders of magnitude and occur in different types of star-planet systems and evolutionary phases, which further affects the distributions of planetary population$^{18,23}$. Despite a large number of studies dedicated to atmospheric escape$^{30-31}$, there is a lack of a conceptual diagram that illustrates the driving mechanism and escaping pattern for low-mass exoplanet population as discovered by Kepler.

\subsection{Classification of Driving Mechanisms by Upgraded Jeans Parameter}

A comprehensive analysis was conducted on a large sample of approximately 2700 planets orbiting G- and M-type stars, considering both XUV heating and the influence of tidal forces (see the sections of The sample, The model and boundary condition, The evolution of XUV emission and Role of X-ray in Methods). It is highly significant if one can determine the driving mechanism by using the intrinsic parameters of the planets and the host stars, without employing sophisticated (numerical) models. One can expect that the Jeans parameter is suitable to make a classification because it gives the ratio of gravitational energy to thermal energy. The range of of $\lambda$ values for these planets spans from less than 1 to $\sim$ 300 (Extended data Fig. 1a). The results can be initially classified into two groups based on the value of $\lambda$. My calculation results show that the TE-driven escape only occurs if Jeans parameter $\lambda$ is smaller than $\sim 3$ (for details see the section of the TE-driven escape in Method and Extended data Fig. 2), which defines the boundary of TE-driven escape and is also consistent with the results of Monte Carlo simulation$^{28}$. Otherwise, as mentioned above, the driving mechanisms of other planets can be either tidally- or XUV-driven, which will be further investigated in the subsequent analysis.

In order to clearly separate the tidally- and XUV-driven escape, I recalculated the atmospheric escape for the planets with 3$<\lambda < $ 40 ($\sim$ 800 model planets) neglecting the stellar tidal forces. The left panel of Fig. 1 shows the ratio of mass loss rates predicted neglecting tidal forces ($\dot{M}_{no tide}$) to those with tidal forces ($\dot{M}_{tide}$) against Jeans parameter. For some planets, the ratios of the mass loss rate are lower than 0.5 and decrease to even 10$^{-4}$ (blue and green asterisks in Fig. 1) while the rest is larger than 0.5 (black and green triangles). The escape is defined as tidally-driven if $\dot{M}_{no tide}/\dot{M}_{tide}<0.5$, otherwise it is XUV-driven. However, it is clear from the left panel of Fig. 1 that the planets can not be classified well with $\lambda$. This means that some essential physics are missed. I stress that any combinations of the mass, radius and temperature of the planets do not reflect the effect of the stellar tidal forces. Thus, by introducing a dimensionless potential energy reduction factor$^{7}$, $\textmd{K}=1-\frac{3}{2f}+\frac{1}{2f^{3}}$ (where f=$\frac{R_{Roche}}{R_{p}}=(\frac{M_{p}}{3M_{s}})^{1/3}\frac{d}{R_{p}}$, $\emph{d}$ is the orbital distance, R$_{Roche}$ is the radius of Roche lobe. M$_{s}$ is the mass of the host star), I constructed an upgraded Jeans parameter to express the reduction of the gravitational potential energy of planet owing to the stellar tidal forces. The upgraded Jeans parameter is defined as $\lambda^{*}$=$\lambda \ast$K (K$<$1). I used it to recheck the properties of the ratios of mass loss rate. As shown in the right panel of Fig. 1, there are three types of aggregation that is differentiated by $\lambda^{*}$. The atmospheric escape of those planets with $\lambda^{*}<3$ is tidally-driven because of $\dot{M}_{no tide}/\dot{M}_{tide}<0.5$, which is also supported by the previous results if the upgraded Jeans parameters are calculated at the same lower boundary altitude (ref.$^{23}$). The rest can be divided into two regimes: tidally-XUV-driven transition regime (3$<\lambda^{*}<$6), in which some of the planets appear tidally-driven while the rest are XUV-driven, and XUV-driven regime ($\lambda^{*}>$6), in which the ratios of $\dot{M}_{no tide}/\dot{M}_{tide}$ are higher than 0.5. My results also demonstrate that for M-type stars, the distributions of the ratios of mass loss rates are nearly identical to those of G-type stars (green asterisks and triangles of Fig. 1).

%\textbf{in which some of the planets appear tidally-driven while others rely on XUV radiation or the planet's own thermal energy as the main energy sources of escape (see the section of the TE-driven escape in Method and Extended data Fig. 2 )}

Fig. 2 demonstrates the dependence of the mass loss rate of planets with different driving mechanisms on F$_{XUV}/\rho$ (where F$_{XUV}$ and $\rho$ are the stellar XUV flux at the planetary orbit and mean densities of the planets, respectively). Most XUV-driven planets (triangles in Fig. 2) with relatively high densities (Extended Data Fig. 1b) follow an approximately linear relationship between the mass loss rates and F$_{XUV}/\rho$, as expected from the energy-limited equation$^{8,40}$, and the outflow of their atmospheres is transonic (black and green triangles of Fig. 2). Planets with high gravitational potential ($\Phi$) deviate significantly from the approximately linear relationship. Therefore, for planets with $\Phi<10^{13}$erg/g, the fit of $\dot{M}$ to F$_{XUV}/\rho$ can be expressed as
\begin{equation}
log \dot{M}=6.83(\pm0.0205)+0.91(\pm0.00485)logF_{xuv}/\rho.
\end{equation}

Compared to the XUV-driven escape, the TE- and tidally-driven escapes exhibit different distributions. These types of escapes occur on exoplanets with a low density (Extended Data Fig. 1b). Their mass loss rates are not related to F$_{XUV}/\rho$, thus indicating different mechanisms from XUV-driven escape. In addition, the velocity of TE-driven escape at the lower boundary of the simulation domain is imposed to be equal to the sound speed, which limits the mass loss rate (see the section of The model and boundary conditions in Methods). In Extended data Fig. 3a, I also show that the velocities of the atmospheres of certain planets remain supersonic throughout the whole domain if the lower boundary velocity is not forced to equal the speed of sound. As a result, the mass loss rates are higher. The classification of the driving mechanisms can also be verified further by comparing the altitudes of Roche radius, the mean absorption radius of XUV photons (R$_{XUV}$) and the location of sonic point (R$_{s}$) (see the left panel of Extended Fig.4 and the section of Distributions of R$_{XUV}$, R$_{Roche}$ and R$_{s}$ in Methods).

\subsection{Neutral and Ionized escaping atmosphere}

In the regime of XUV driven escape, my results show that the degree of ionization is a function of $\lambda^{*}$ and gravitational potential. I defined that the atmosphere is neutral in the partially ionized gas if the altitude of sonic point is lower than the location of transition of H/H$^{+}$ (where the number density of H$^{+}$ exceeds the number density of H). Otherwise, the atmosphere is ionized and dominated by H$^{+}$. Fig. 3 shows that if the XUV radiation is lower, the more planets will be in regime of neutral atmosphere. The left panel of Fig. 3 shows that the regimes of ionized and neutral atmospheres does not explicitly depend on $\lambda^{*}$, and the slope of the boundary between neutral and ionized atmospheres seems to have a turning point at the locations of $\lambda^{*}\sim 70-80$ and F$_{\mathbf{xuv}}\approx6$$\times$10$^{4} erg/cm^{2}/s$. In fact, the degree of ionization is a balance between ionization and recombination (the process of recombination is inversely related to the temperature), which is determined by a combination of XUV irradiation, the density and temperature structures. Moreover, the density and temperature profiles are very sensitive to the gravitational potential$^{41}$. Thus, I also show the dependence of the ionization degree on gravitational potential in the right panel of Fig. 3. There is a clear boundary that divides the ionized and neutral atmospheres. With the decline of F$_{\mathbf{xuv}}$, the boundary between ionized and neutral atmospheres moves to $\Phi=\sim 8.7\times10^{12}$ erg/g. For the case of small gravitational potential their density profiles are flatter, which means that the penetration of photons encounters a barrier owing to the high number density of hydrogen atoms at high altitudes. This explain why the ionized winds only occur on the regimes of $\Phi > \sim 10^{12}$ erg/g even if the XUV irradiation is in a very high level. In the case of low XUV radiation, the photoionization is relatively weak so that only those planets with high gravitational potential ($\Phi >8\times10^{12}$erg/g) appear ionized winds because of the rarefied number density and higher temperature (In Extended data Fig. 5, I show the atmosphere profiles of two planets with different gravitational potentials to demonstrate the difference of the density and temperature around the turning point of neutral and ionized winds). Finally, the altitude of transition of H/H$^{+}$ can be related to the transition of XUV-driven escape among energy-, photo- and recombination-limited regimes$^{42}$ (for details see the section of The energy-, photo- and recombination-limited regimes in Methods).

\subsection{Subsonic Escape}
In the XUV-driven regime, with a decrease of XUV irradiation and increase of gravitational potential, the XUV energy drives a weak wind and the stellar photons only produce sparse ions. As a consequence of infrequent collisions of particles, the mean free path can be greater than the scale height of the flow so that the sonic point is higher than the location of exobase. Such planetary wind is defined as subsonic escape or slow hydrodynamic escape$^{43-45}$, which is in an intermediate case between the Jeans escape and the transonic hydrodynamic escape. For instance, the slow hydrodynamic escape can occur for Pluto if the value of Jeans parameter at the position of 1.2 R$_{Pluto}$ is around 20$^{44-45}$. On the other hand, from the point of view of energy, the XUV energy deposited in the atmosphere can not drive a transonic outflow if the net energy absorbed from the XUV radiation is smaller than the critical value Q$_{c}$ $^{46}$ (for details see the section of The criterion of supersonic and subsonic escape in Methods). In the neutral atmospheres, a subsonic hydrodynamic escape can occur (filled black triangles in Fig. 3) in the regime of low XUV irradiation and high gravitational potential ($5\times 10^{12}<\Phi<10^{13}$ erg/g). For planets with ionized atmosphere, a relatively low number density of ions can fulfill the condition of transonic escape owing to the smaller mean free path of ion collisions (H$^{+}$-H$^{+}$). As expected, the outflows of ionized atmosphere are always transonic. It is clear from Fig. 2 that for neutral subsonic escape (red triangles), the mass loss rates are lower than those of most transonic escapes though they lie at the upper boundary of the distribution of the mass loss rates  for a given F$_{XUV}/\rho$. The upper limit is around 4$\times$10$^{8}$ g/s while the lowest mass loss is about 6$\times$ 10$^{6}$ g/s. The value of 6$\times$ 10$^{6}$ g/s is comparable to the predicted mass loss rate of Titan in a transonic flow (the XUV radiation is 400 times the current value). In addition, the atmospheric escape of Titan converts to subsonic hydrodynamic escape if the EUV flux is 100 times the current value$^{47}$. Fig. 2 also shows that the Jeans escape rates of the planets in solar system are much lower than those of their hydrodynamic escape owing to their lower equilibrium temperatures. For close-in planets of subsonic group in my sample, the range of their Jeans parameters at R$_{p}$ are from $\sim$60 to 200. The atmospheres will be very compact (R$_{p}\sim$R$_{exobase}$) if the XUV heating is neglected. Thus, the Jeans's escape rate is much smaller than the hydrodynamic escape rate. For the planets with very high gravitational potentials, there is the numerical oscillations for the velocity distribution. The oscillation is caused by numerical inaccuracy$^{24}$ and can be compensated at some extent by increasing the resolution of grid. Our results show that the mass loss rates of the planets with very high gravitational potentials are in the order of magnitude of 10$^5$ g/s (the change due to the oscillations does not exceed the factor of a few), which hints that the escape of these planets could occur in the intermediate regime between hydrodynamic and Jeans escape.

\subsection{Discussion}
In the past several decades, Jeans parameter has been a good indicator for distinguishing the escaping mechanism, such as the Jeans and hydrodynamic escape. As a significant generalization of Jeans parameter in the regimes of close-in exoplanets, my results show that the upgraded Jeans parameter is also analogous to the case without the tidal forces. For example, for Mars-like planets with a larger orbital distance ($\sim$1 AU), $\lambda=2.5$ was found to be a critical value for a transition from XUV-driven stationary hydrodynamic escape to TE-driven atmospheric escape$^{48}$. At the separation of $\sim$1 AU, the tidal forces can be neglected so that $\lambda^{*}\approx\lambda$. The value of 2.5 is similar to my critical value. Contrary to the study of the planets in solar system, the Jeans escape does not occur for close-in exoplanets while the value of $\lambda^{*}$ is larger than 3. In fact, the hydrodynamic escape can occur even if the value of $\lambda^{*}$ is beyond 100 (Extended data Fig. 1a), which is caused by XUV heating.

In the case of XUV-driven escape, the energy-limited equation is convenient for obtaining the mass loss rate of XUV-driven escape. However, the heating efficiency and the effective XUV absorption radius are not known. To facilitate using the energy-limited formula, I constrain the product$^{31}$ of the heating efficiency ($\eta$) and the square of the ratio of XUV mean absorption radius to planetary radius ($\beta$=R$_{xuv}$/R$_{p}$) by equating the results of hydrodynamic simulations to the predictions of the energy-limited equation, namely, $\eta\beta^{2}=\frac{\dot{M}_{hydro} G M_{p} K}{\pi F_{XUV} R_{p}^{3}}$. The resulting distributions of log$\eta\beta^{2}$ against XUV and revised gravitational potential ($\Phi'$=$\Phi$*K) are shown in Fig.4. In general, the heating efficiency increases with the gravitational potential and decreases after $log\Phi> 13$ while R$_{xuv}$ of planets with lower gravitational potential is larger than those of high gravitational potential$^{31,49}$. $\eta\beta^{2}$ is lower than 0.01 if their revised gravitational potentials are higher than $10^{13}$ erg/g (left panel). Such behavior is due to the fact that R$_{xuv}$ is closer to the planetary surface, and the heating efficiency is very low because the high temperature ($\sim$10000K) atmospheres radiate most of the XUV radiation via the strong cooling of Ly$\alpha$$^{5}$. For those planets bathed in high XUV radiation, most of the XUV radiation is used to ionize gas (Fig. 3) and the radiative cooling is also strong$^{5}$. As a consequence, $\eta\beta^{2}$ is small owing to the low heating efficiency and small R$_{xuv}$ (R$_{xuv}$ is close to the planetary surface due to the high ionization degree). For planets in moderate gravitational potential (12$<log\Phi'<13$ erg/g) and relatively low XUV radiation ($F_{XUV}<10^{5} erg/cm^{2}/s$), the relatively high heating efficiency and moderate R$_{xuv}$ together result in the maximum of $\eta\beta^{2}$ (Yellow and Green). In the bottom left corner, a local maximum of $\eta\beta^{2}$ is caused mainly by large R$_{xuv}$. The calculations are conducted in a smaller parameter space for planets orbiting M-type star (right panel). The values of log$\eta\beta^{2}$ are around $\sim$0.1 and slightly increase with $\Phi'$ at a given F$_{XUV}$.

The results above have a few potential applications. First, in order to predict the driving mechanisms for the atmospheric escape, one only needs to know the intrinsic physical parameter of the planet and star. Second, in the cases of TE- and tidally-driven escape the mass loss rates are in the order of magnitude of 10$^{13}$-10$^{15}$ g/s. One can expect that in this situation, the exoplanets could be evaporated to bare cores because a planet with a few earth mass can loss their H-He envelopes in 10$^{7}$-10$^{8}$ years. Such situation can occur in the extremely early phase of planet formation. Third, the calculation results provide a way to fit the uncertain parameters of the energy-limited equation, which is useful for obtaining the realistic mass loss rates.

\begin{figure}
\begin{minipage}[t]{0.5\linewidth}
\centering
\includegraphics[width=3.6in,height=2.5in]{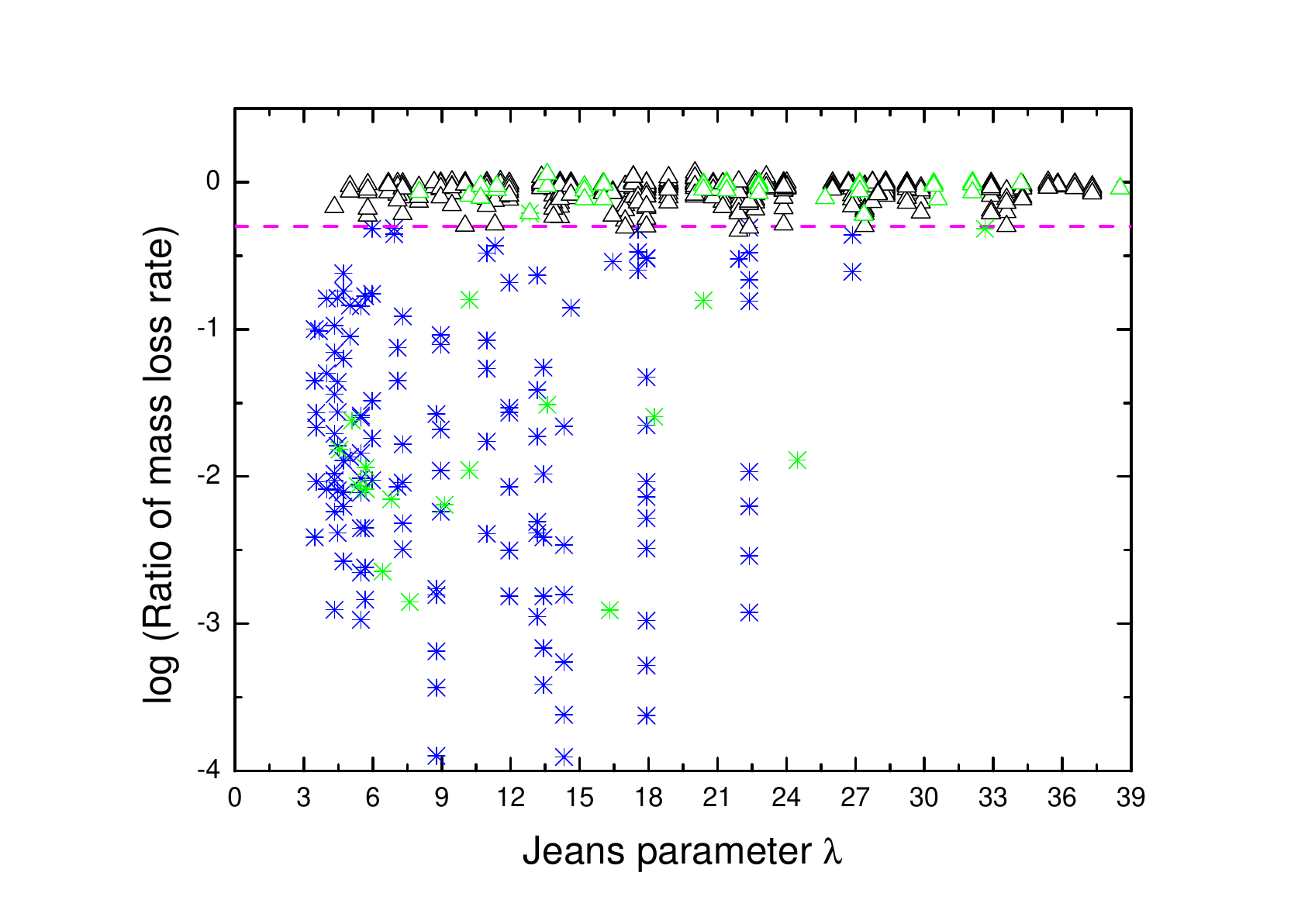}
\end{minipage}
\begin{minipage}[t]{0.5\linewidth}
\centering
\includegraphics[width=3.6in,height=2.5in]{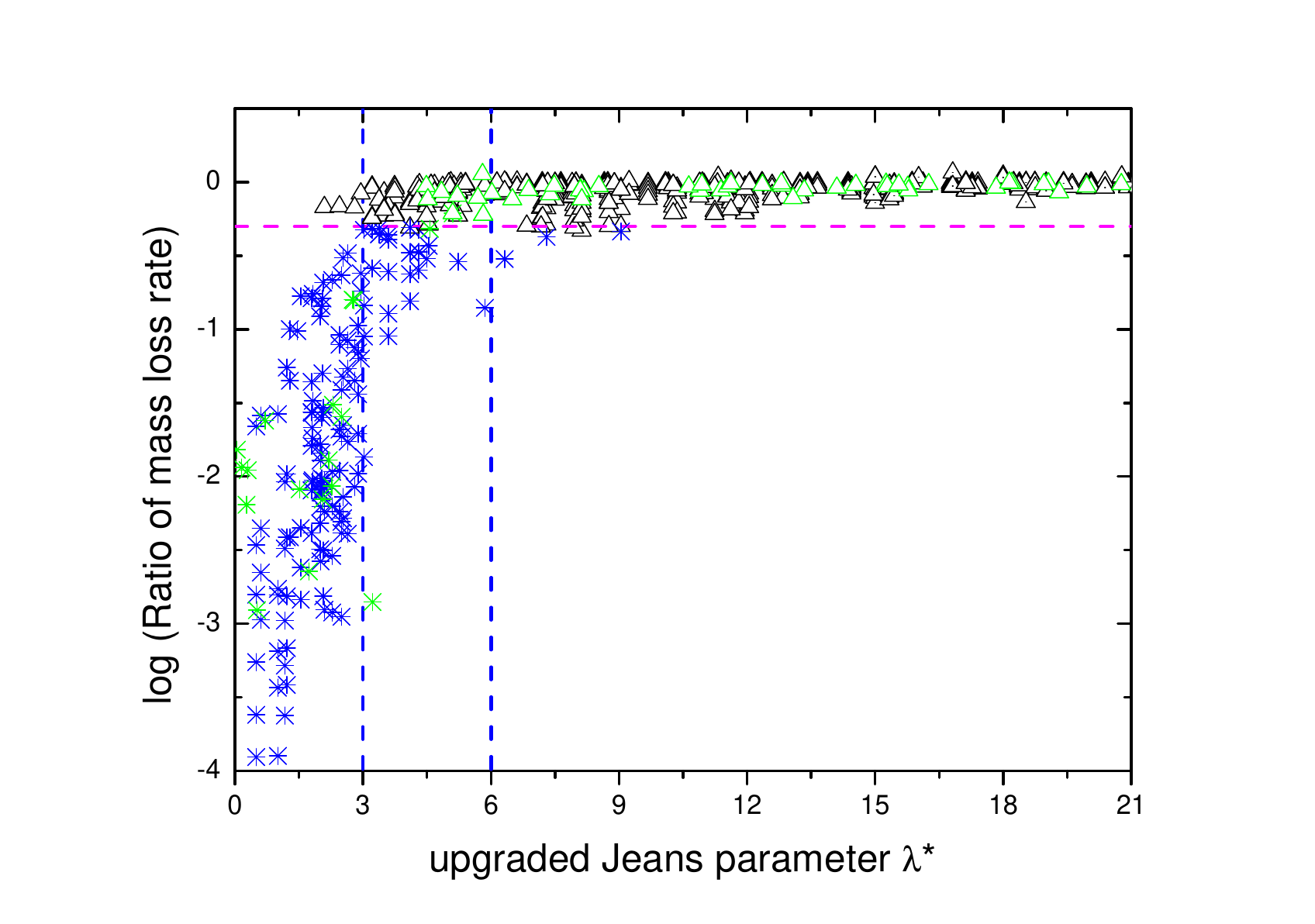}
\end{minipage}

\caption*{\textbf{Fig. 1$\mid$The ratios of mass loss rates without and with tidal forces.} For planets orbiting G-type star, Blue``asterisk'': tidally-driven escape. ``Black triangle'': XUV-driven escape. I also included the planets around M-type stars. Their symbols are the same as those orbiting G-type stars but the color is green. The left panel shows the case of Jeans parameter while the right panel shows the classifications by upgraded Jeans parameter. The left ($\lambda^{*}=3$) and right ($\lambda^{*}=6$) vertical dashed lines in right panel separate the tidally-driven, tidally-XUV-driven transition, and XUV-driven regimes. The horizontal magenta dashed lines in both left and right panels denote the case of $\dot{M}_{notide}$/$\dot{M}_{tide}$=0.5.}
%\label{fidu_trans}
\end{figure}

\begin{figure}[tb]
\centering

\begin{subfigure}[t]{1.0\textwidth}
    \includegraphics[width=\linewidth,valign=t]{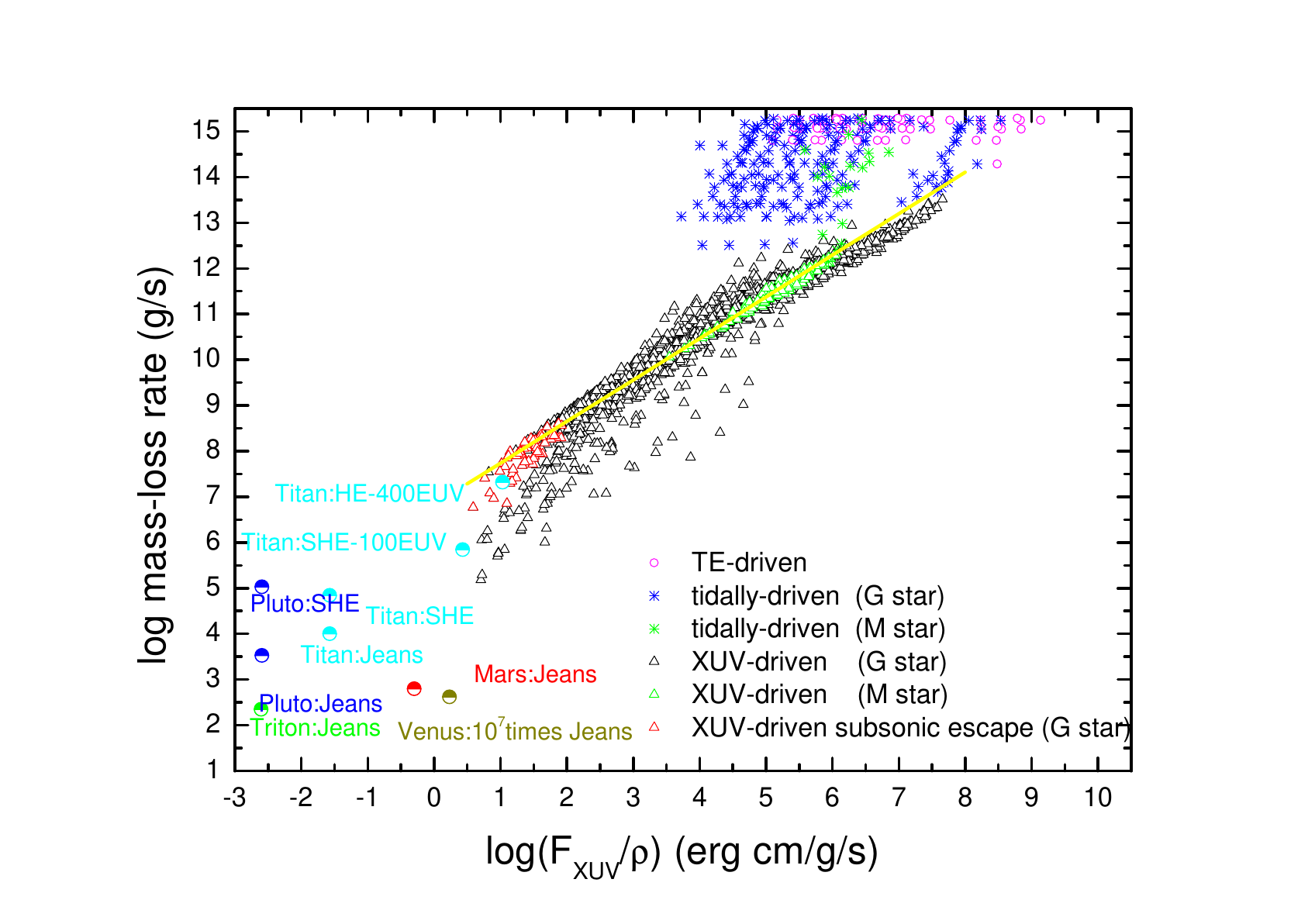}
%    \includegraphics{mass loss rate_F.pdf}
%    \label{Fig2a}
\end{subfigure}\hfill

\caption*{\textbf{Fig. 2$\mid$The mass loss rates of planets at different F$_{XUV}$/$\rho$.} `Magenta circle'': the TE-driven. For planets orbiting G-type star, ` Blue``asterisk'': tidally-driven escape.  ``Black triangle'': XUV-driven escape. ``Red triangle'': XUV-driven subsonic escape in the neutral atmosphere. The subsonic escape is a slow hydrodynamic escape (SHE)  for which the sonic points are higher than the locations of exobase. I also included the planets around M-type stars. Their symbols are the same as those orbiting G-type stars but the color is green. The yellow line denotes the approximately linear relationship of log$\dot{M}$ and logF$_{XUV}$/$\rho$ (Equation (2)). The sparse region above the XUV-driven group is caused by the XUV radiation difference between 0.05 Gyr and 0.5Gyr. The mass loss rates of the planets in solar system are obtained from the references 44 (Pluto: SHE), 47 (Titan: 400 and 100 times EUV), 67( Titan: SHE), 68 (Titan: Jeans), 69 (Pluto: Jeans), 70 and 71 (Triton: Jeans), 72(Mars and Venus: Jeans). Note that the Jeans escape is the escape rates of hydrogen except Pluto. For Pluto the Jeans escape rate is the sum of N$_{2}$ and CH$_{4}$. }
\label{fidu_trans}
\end{figure}

\begin{figure}
\begin{minipage}[t]{0.5\linewidth}
\centering
\includegraphics[width=3.6in,height=2.5in]{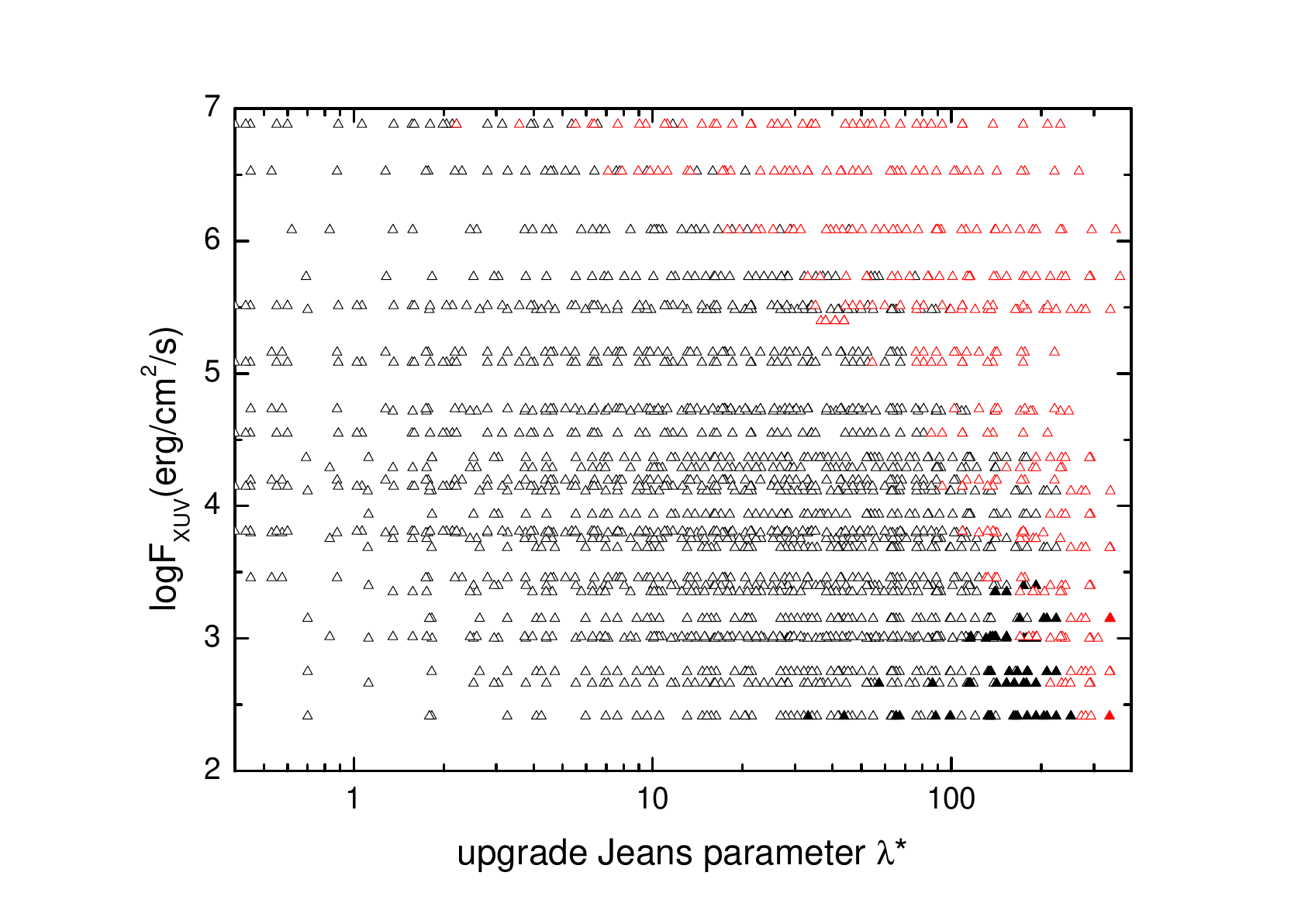}
\end{minipage}
\begin{minipage}[t]{0.5\linewidth}
\centering
\includegraphics[width=3.6in,height=2.5in]{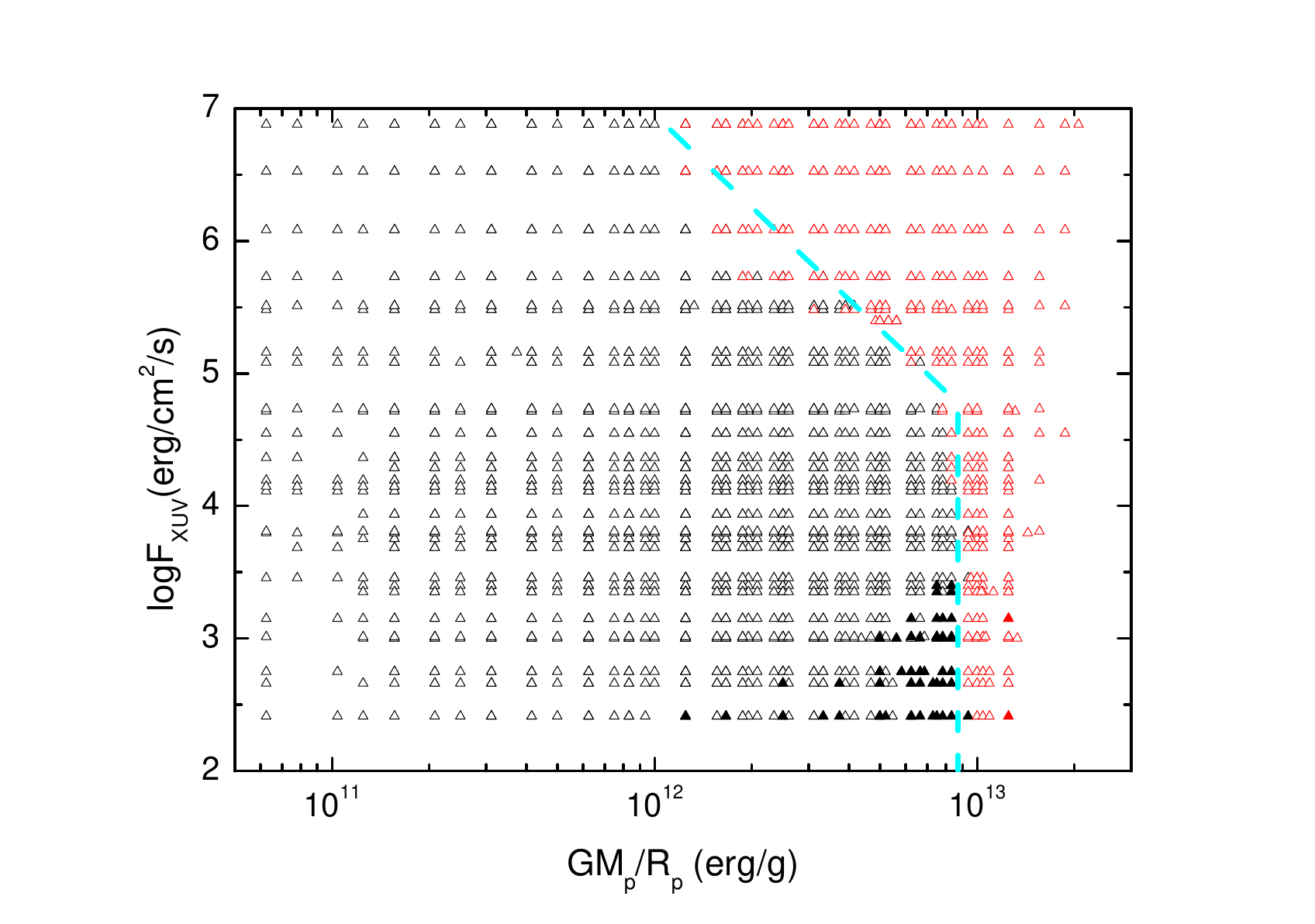}
\end{minipage}

\caption*{\textbf{Fig. 3$\mid$The neutral and ionized atmosphere.} The left and right panels show the dependence of ionization degree on the $\lambda^{*}$ and GM$_{p}$/R$_{p}$, respectively. Black and red denote the neutral and ionized atmosphere and the filled black and red triangles represent the subsonic escape in the neutral and ionized atmospheres, respectively. The light blue dashed lines of the right panel schematically separate the neutral and ionized winds.}
%\label{fidu_trans}
\end{figure}

\begin{figure}
\begin{minipage}[t]{0.5\linewidth}
\centering
\includegraphics[width=3.6in,height=2.5in]{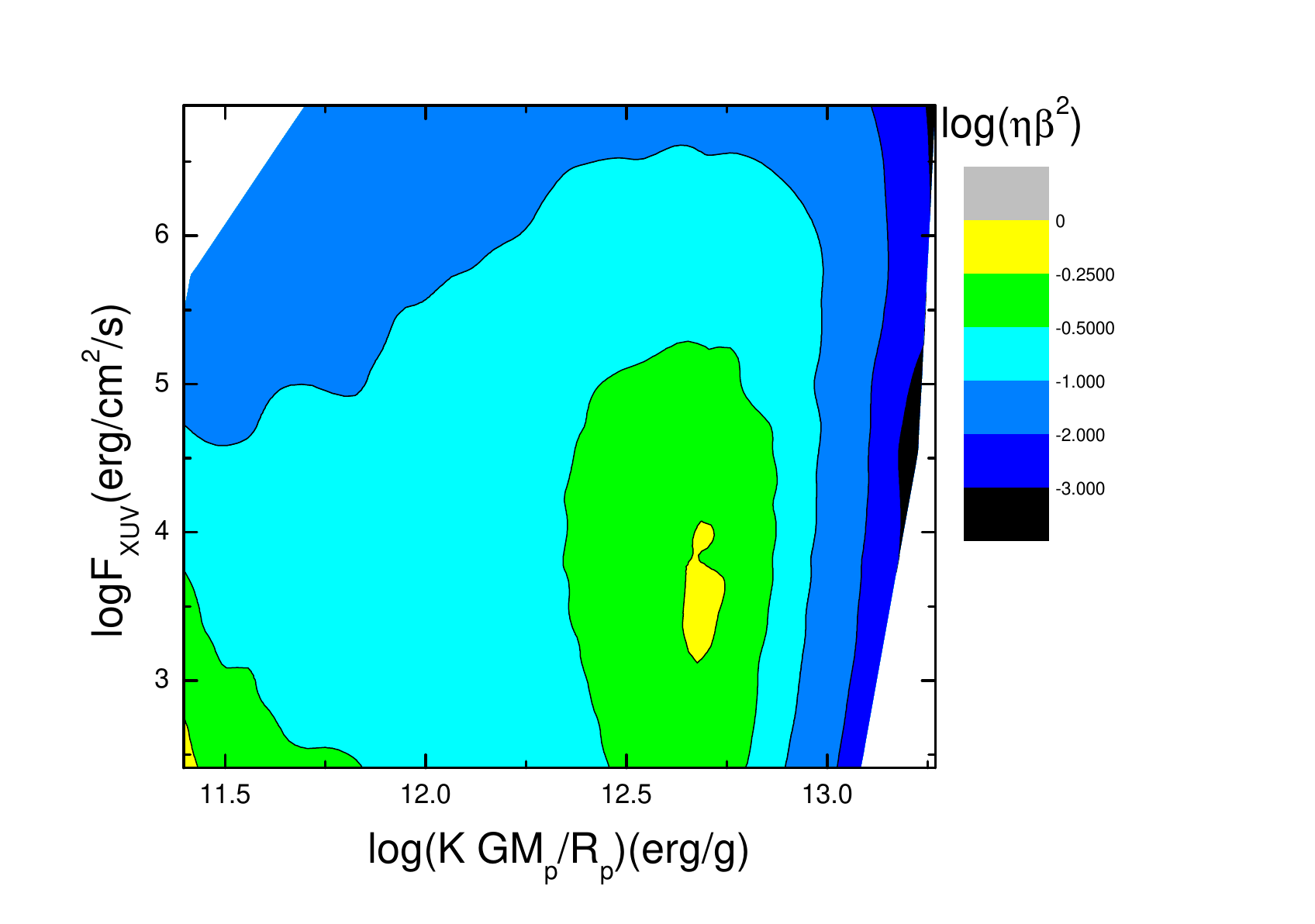}
\end{minipage}
\begin{minipage}[t]{0.5\linewidth}
\centering
\includegraphics[width=3.6in,height=2.5in]{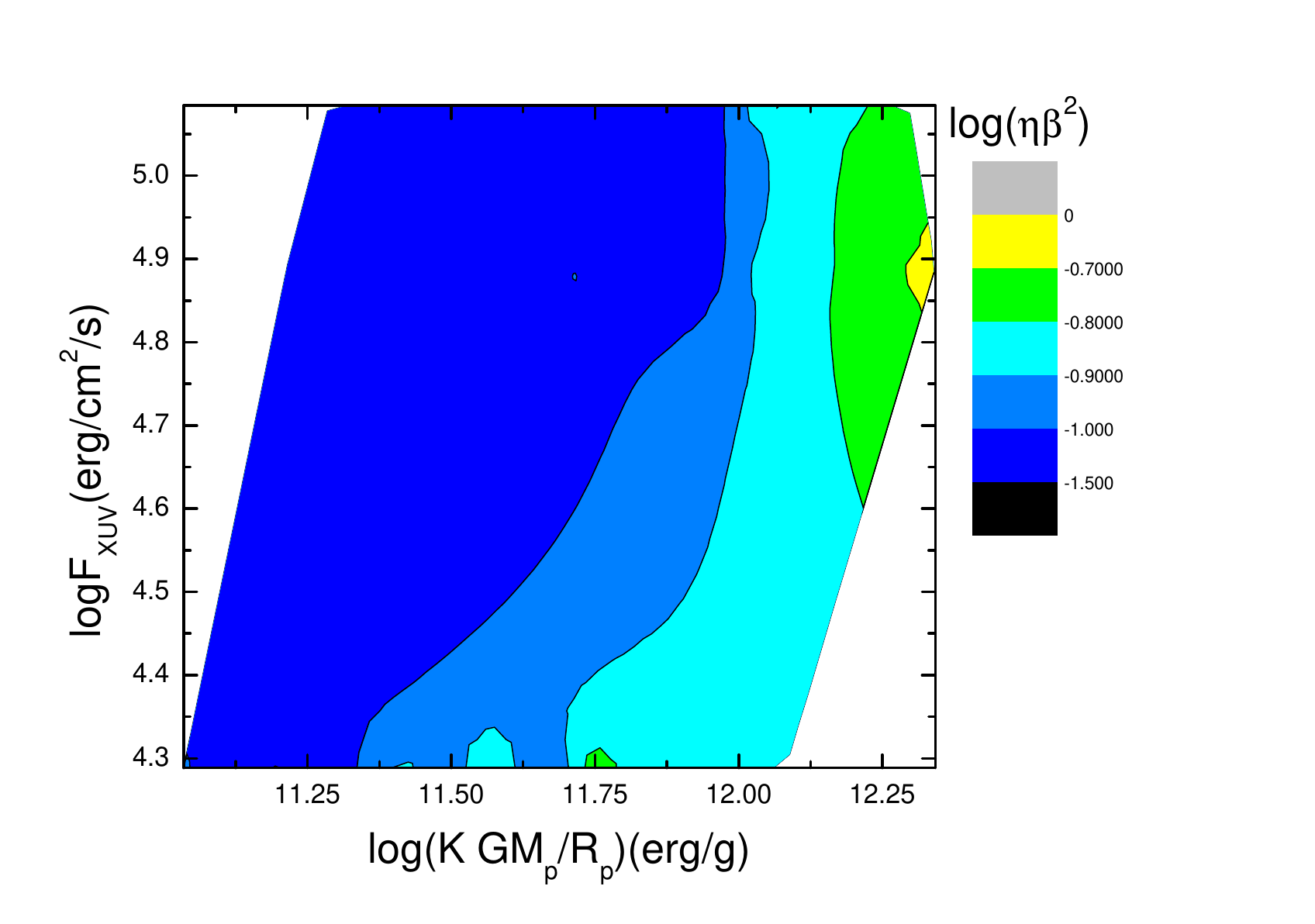}
\end{minipage}
\caption*{\textbf{Fig. 4$\mid$The product of the heating efficiency ($\eta$) and (R$_{xuv}/R_{p})^{2}$.}  The product was obtained by equating the hydrodynamic results with energy-limited formula. Left panel: planets orbiting G-type stars. Right panel: planets orbiting M-type stars.}
\label{fidu_trans}
\end{figure}

\begin{figure}
%\begin{minipage}[t]{0.5\linewidth}
%\centering
%\includegraphics[width=3.0in,height=2.3in]{extend-fig1a.pdf}
%\end{minipage}
%\begin{minipage}[t]{0.5\linewidth}
%\centering
%\includegraphics[width=3.0in,height=2.3in]{extend-fig1b.pdf}
%\end{minipage}
\begin{minipage}[t]{0.5\linewidth}
\centering
\includegraphics[width=3.0in,height=2.3in]{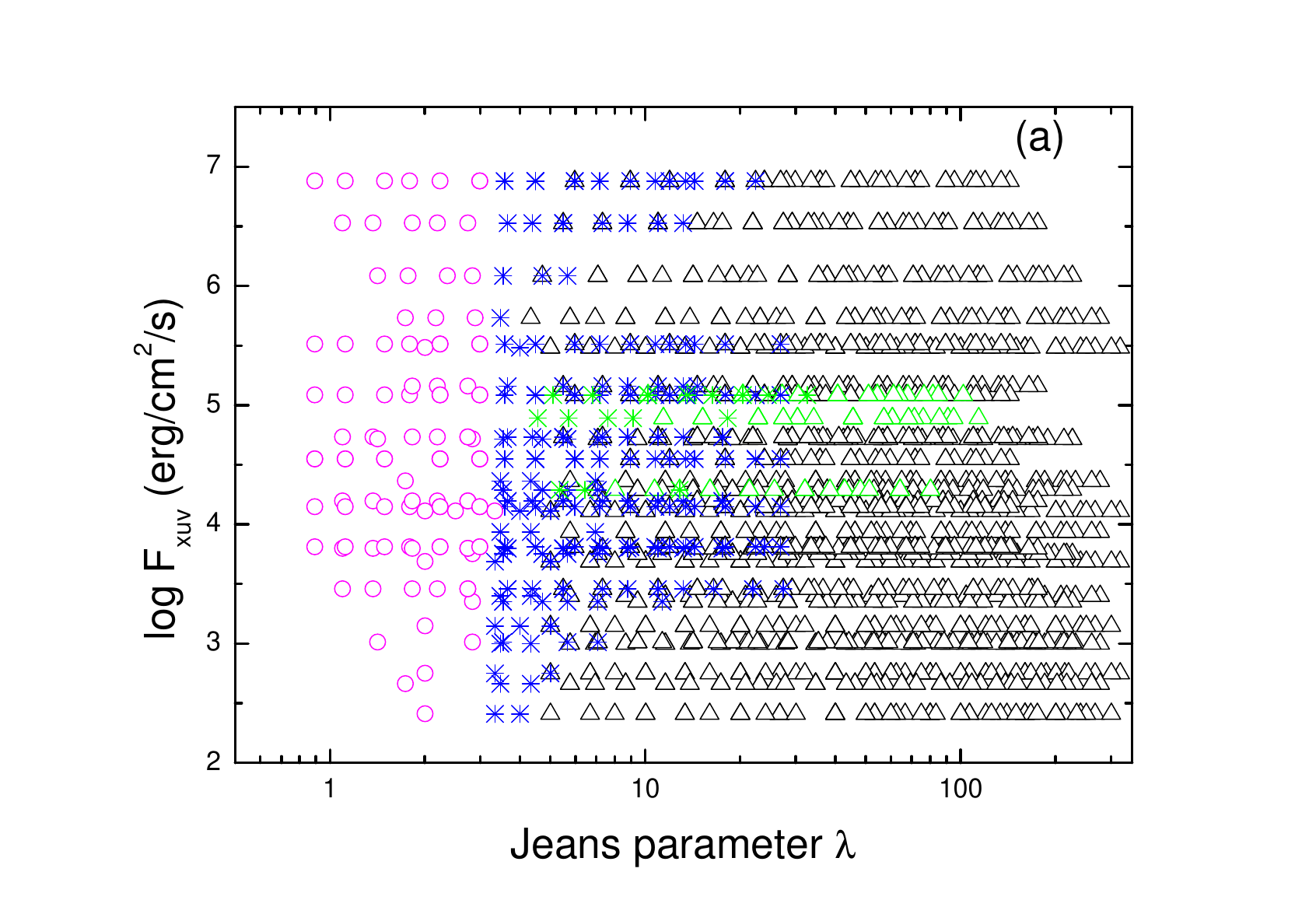}
\end{minipage}
\begin{minipage}[t]{0.5\linewidth}
\centering
\includegraphics[width=3.0in,height=2.3in]{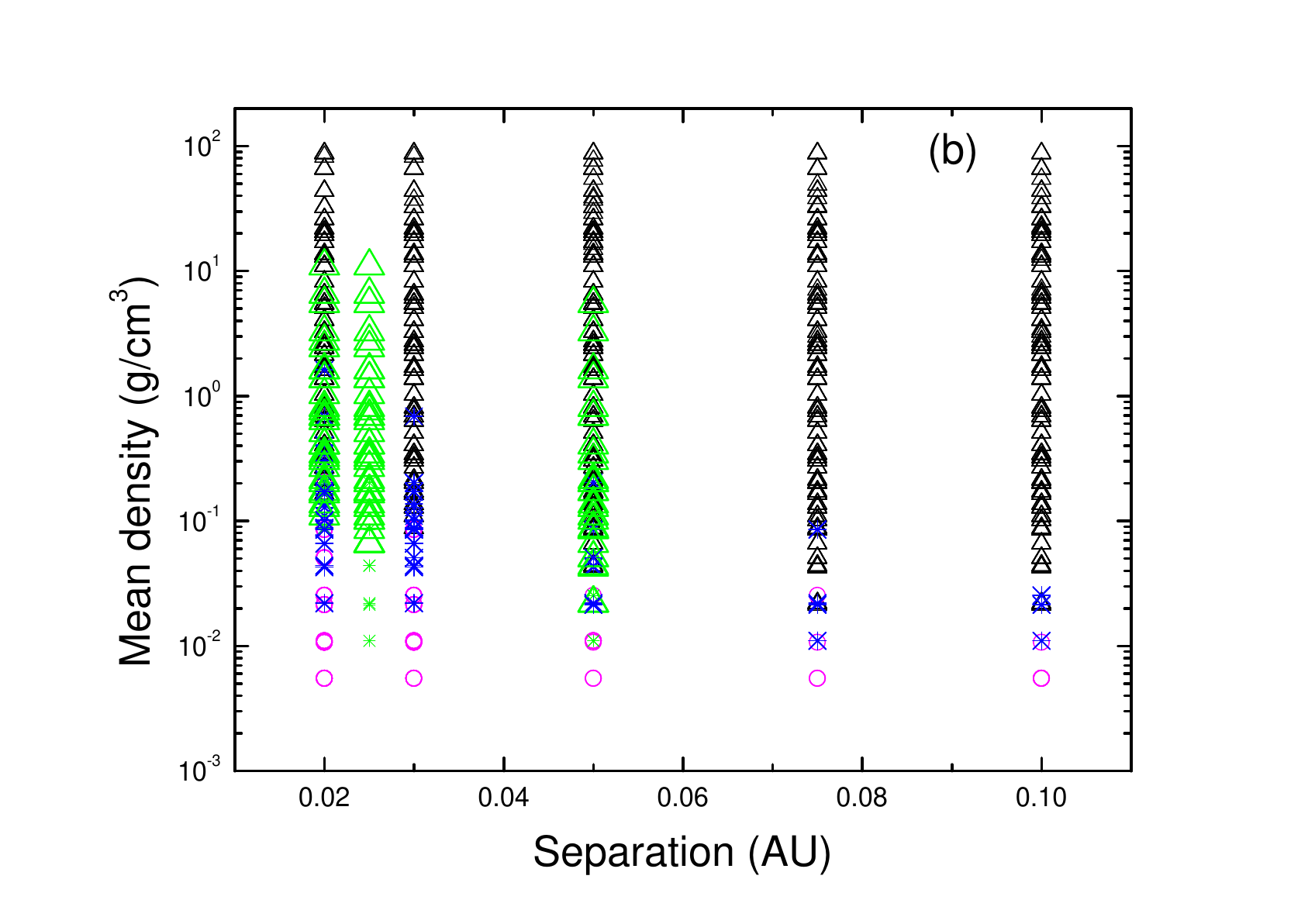}
\end{minipage}

\caption*{\textbf{Extended Data Fig. 1$\mid$ The distributions of density, separation and $\lambda*$ for planets with different driving mechanisms.} The symbols are the same as those in Figure 2. (a) the distributions of $\lambda$ and XUV flux; (b) the distributions of separation and mean density. In generally, the planets have smaller density if their hydrodynamic escapes are TE- and tidally-driven. In contrast, the trend is contrary to the planets with XUV-driven. For TE-driven escape, their $\lambda$ values are smaller than 3. For tidally-driven escape, the values of $\lambda^{*}$ is smaller than 3. In the regime of tidally-XUV-driven transition regime, the values of $\lambda^{*}$ are in the range of 3-6. The values of $\lambda^{*}$ are greater than 6 in XUV-driven regime (see Fig.1).}
\end{figure}

\begin{figure}[tb]
\centering

\begin{subfigure}[t]{1.0\textwidth}
    \includegraphics[width=\linewidth,valign=t]{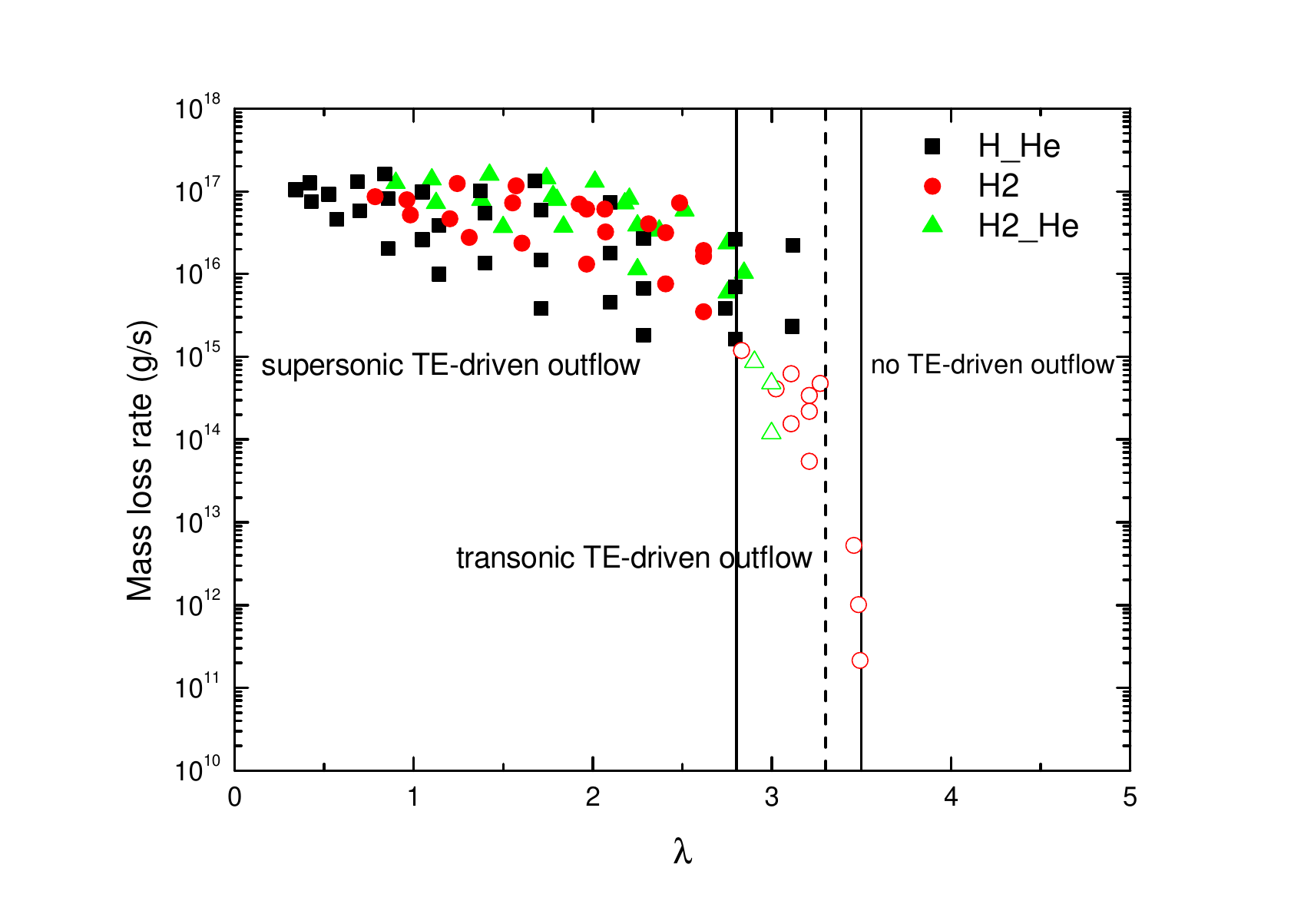}
%    \includegraphics{mass loss rate_F.pdf}
%    \label{Fig2a}
\end{subfigure}\hfill

\caption*{\textbf{Extended Data Fig. 2$\mid$ The TE-driven escape of the atmosphere.} ``Black square'': the atmosphere is composed of hydrogen atoms and Helium; ``Red circle'': the atmosphere of Hydrogen molecular; ``Green triangle'': the atmosphere is composed of hydrogen molecules and Helium. The two solid lines are $\lambda$=2.8 and 3.5, and the region between two solid lines represents the regime of transonic escape for the atmospheres of H$_{2}$ and H$_{2}$+He. The dashed line is $\lambda$=3.3 that denote the the limit of supersonic escape for mon-atomic gases. All filled symbols are supersonic outflow while the triangles and circles denote the case of transonic escape. Note that the velocities at the lower boundary are not imposed to equal with the sound speed such that my model predicts very high the mass loss rates. Also note that the tidal forces, XUV irradiation and all chemical reactions are ignored for all planets in the figure.}
\label{fidu_trans}
\end{figure}

\begin{figure}
\begin{minipage}[t]{0.5\linewidth}
\centering
\includegraphics[width=3.0in,height=2.3in]{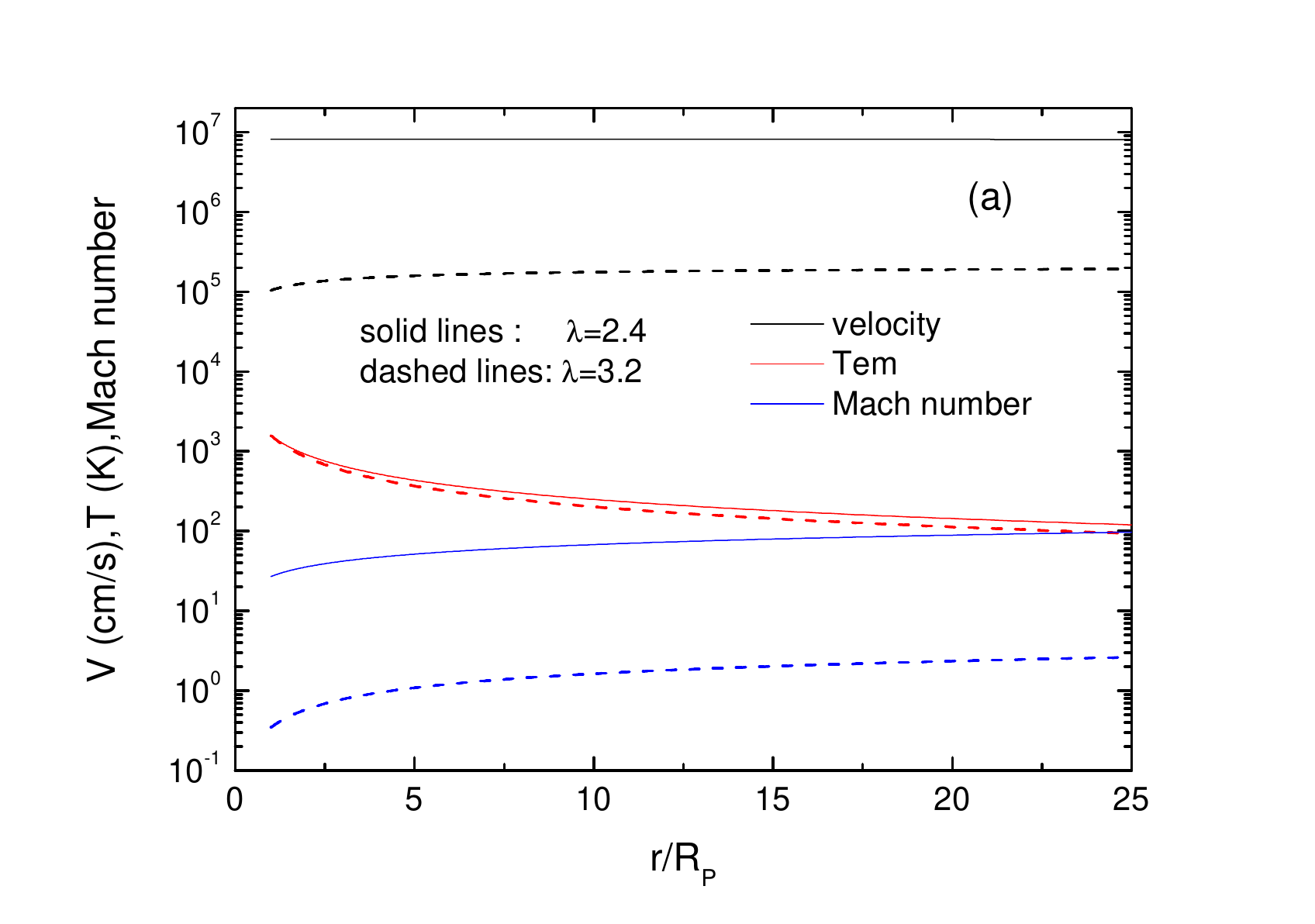}
\end{minipage}
\begin{minipage}[t]{0.5\linewidth}
\centering
\includegraphics[width=3.0in,height=2.3in]{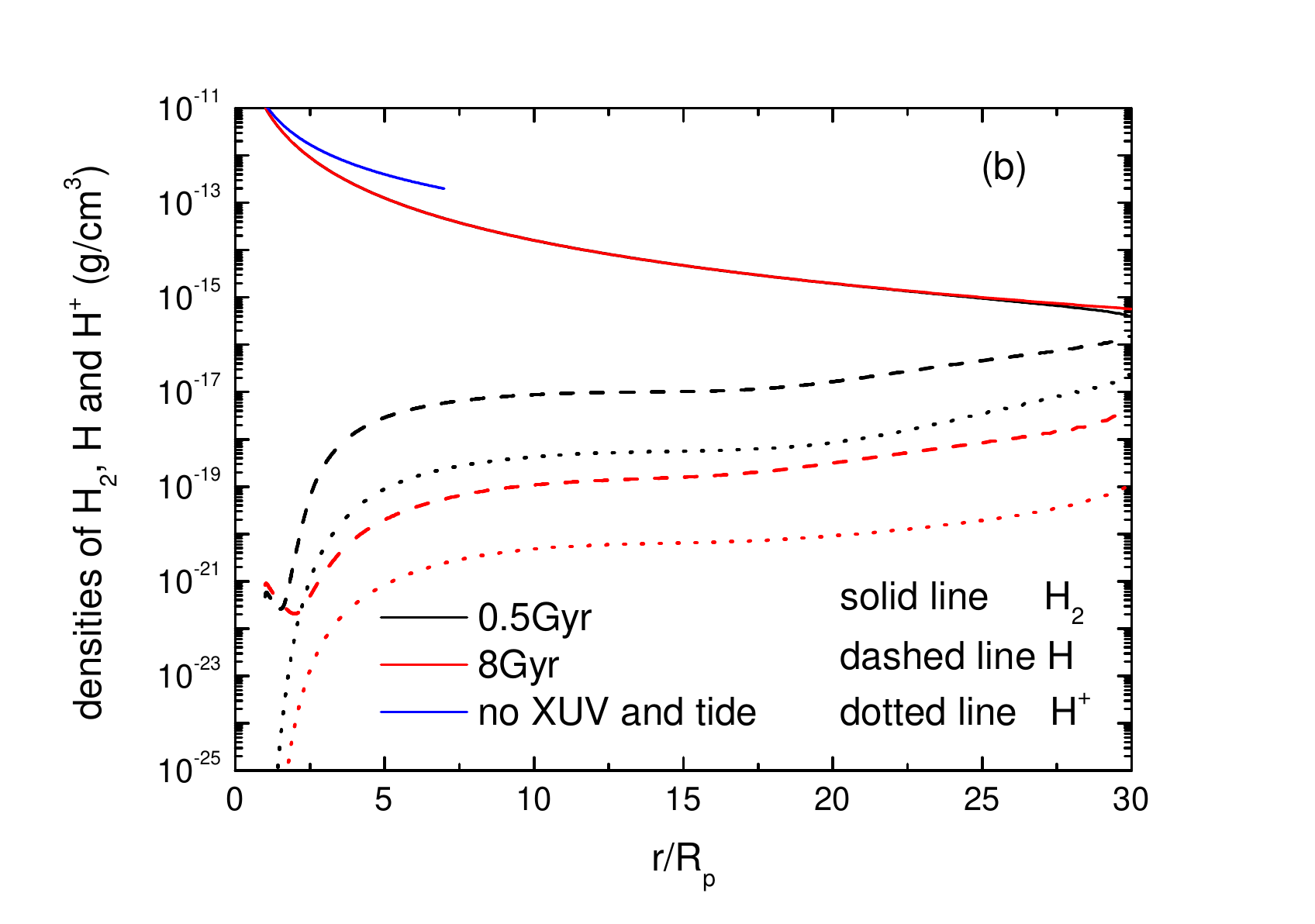}
\end{minipage}
\begin{minipage}[t]{0.5\linewidth}
\centering
\includegraphics[width=3.0in,height=2.3in]{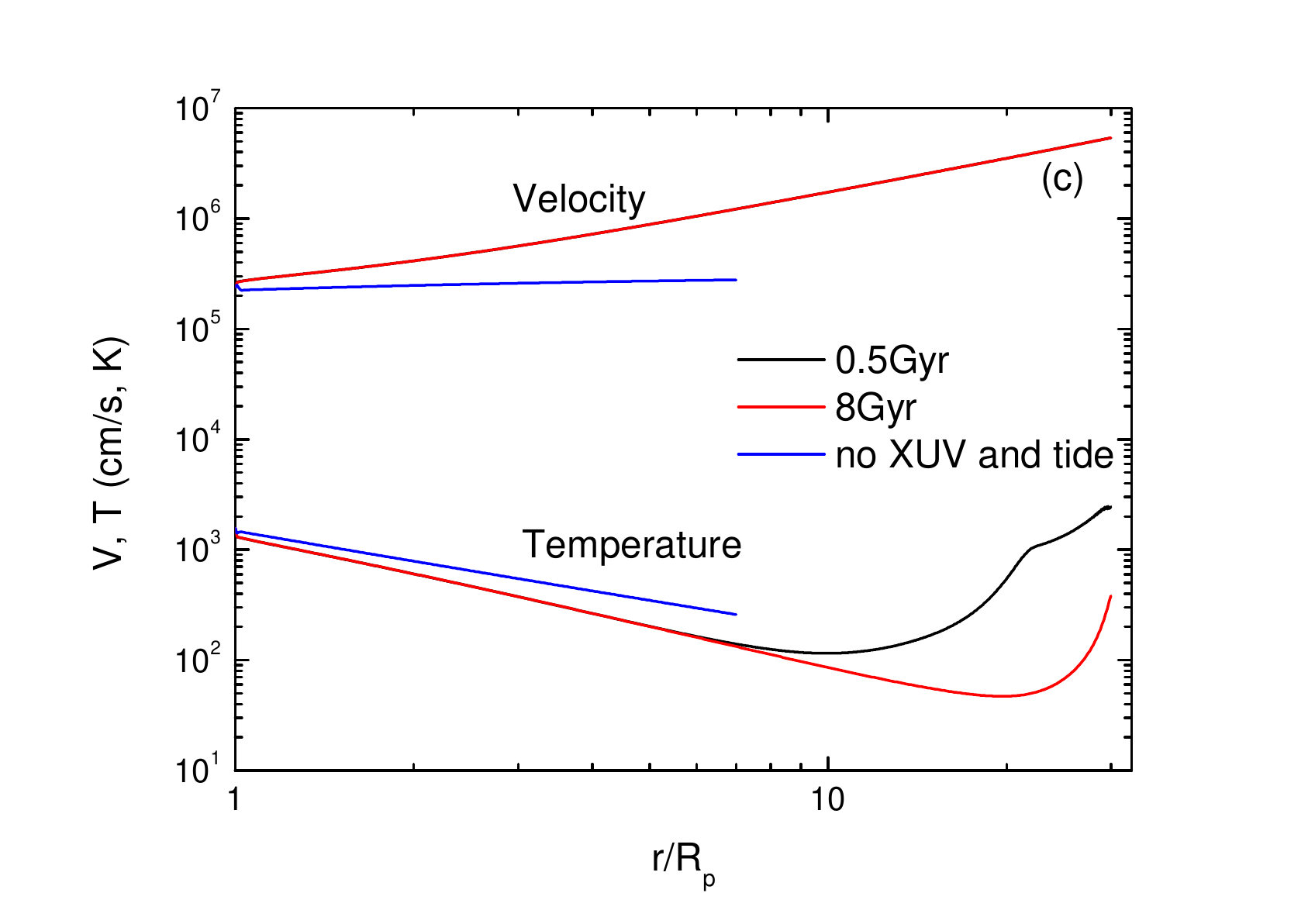}
\end{minipage}

\caption*{\textbf{Extended Data Fig. 3$\mid$ The atmospheric structures of the TE-driven escape.}  In panel (a) I show two samples with different values of $\lambda$. The escape of both the two planets are driven by their thermal energy, but one is supersonic and the other is transonic. Panel(b) and (c) show the number densities of H$_{2}$, H and H$^{+}$ and the profiles of the velocities and temperatures of a planet with M$_{p}=1M_{\oplus}$, R$_{p}=4R_{\oplus}$. I show the cases with 0.5Gyr, 8.0Gyr and no XUV radiation and tidal forces. Note that the case of “no XUV and tide” does not include chemical reactions and should be discriminated from the open circles in Extended data figures 1 and 4.}
%\label{fidu_trans}
\end{figure}

\begin{figure}
\begin{minipage}[t]{0.5\linewidth}
\centering
\includegraphics[width=3.6in,height=2.5in]{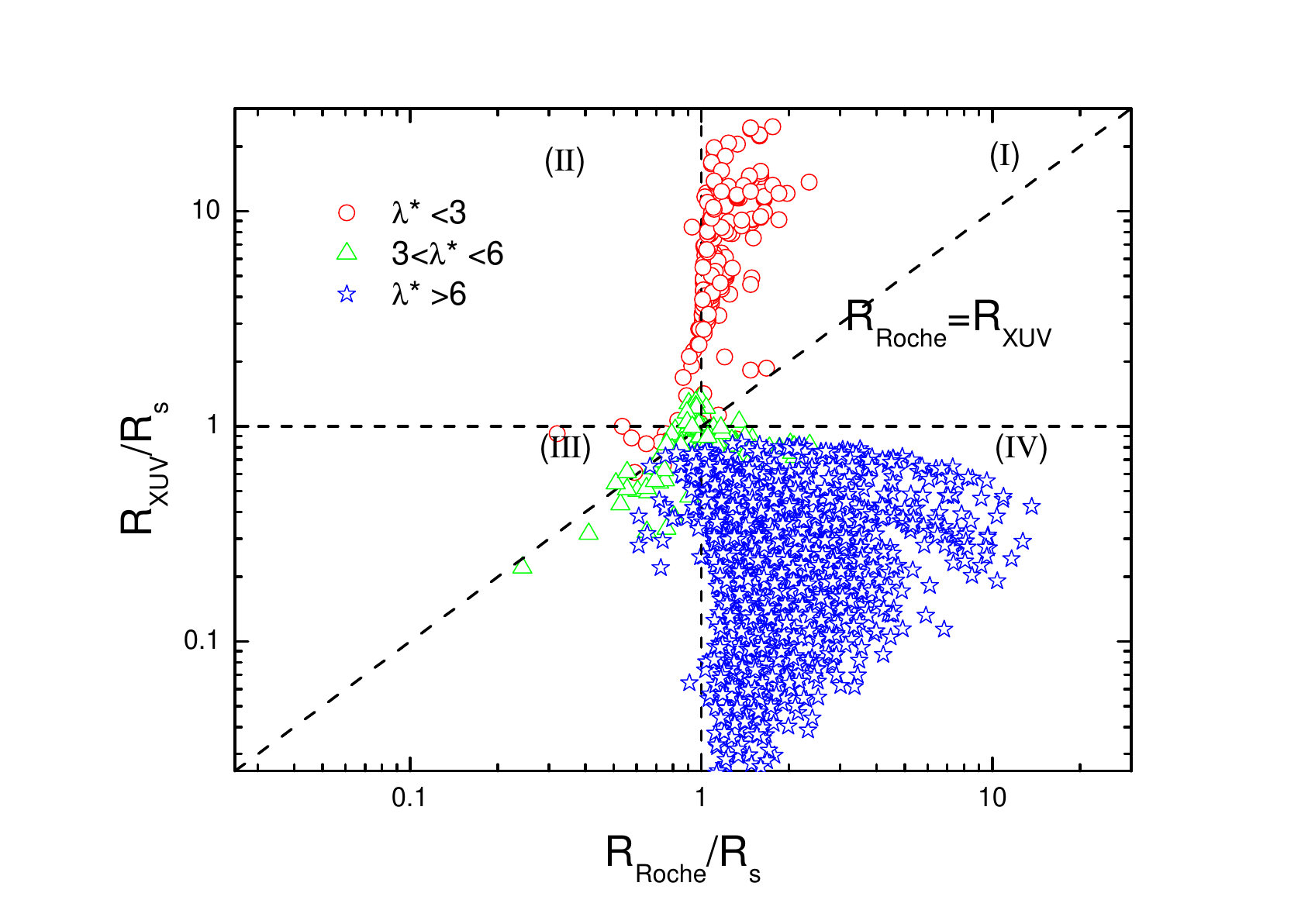}
\end{minipage}
\begin{minipage}[t]{0.5\linewidth}
\centering
\includegraphics[width=3.6in,height=2.5in]{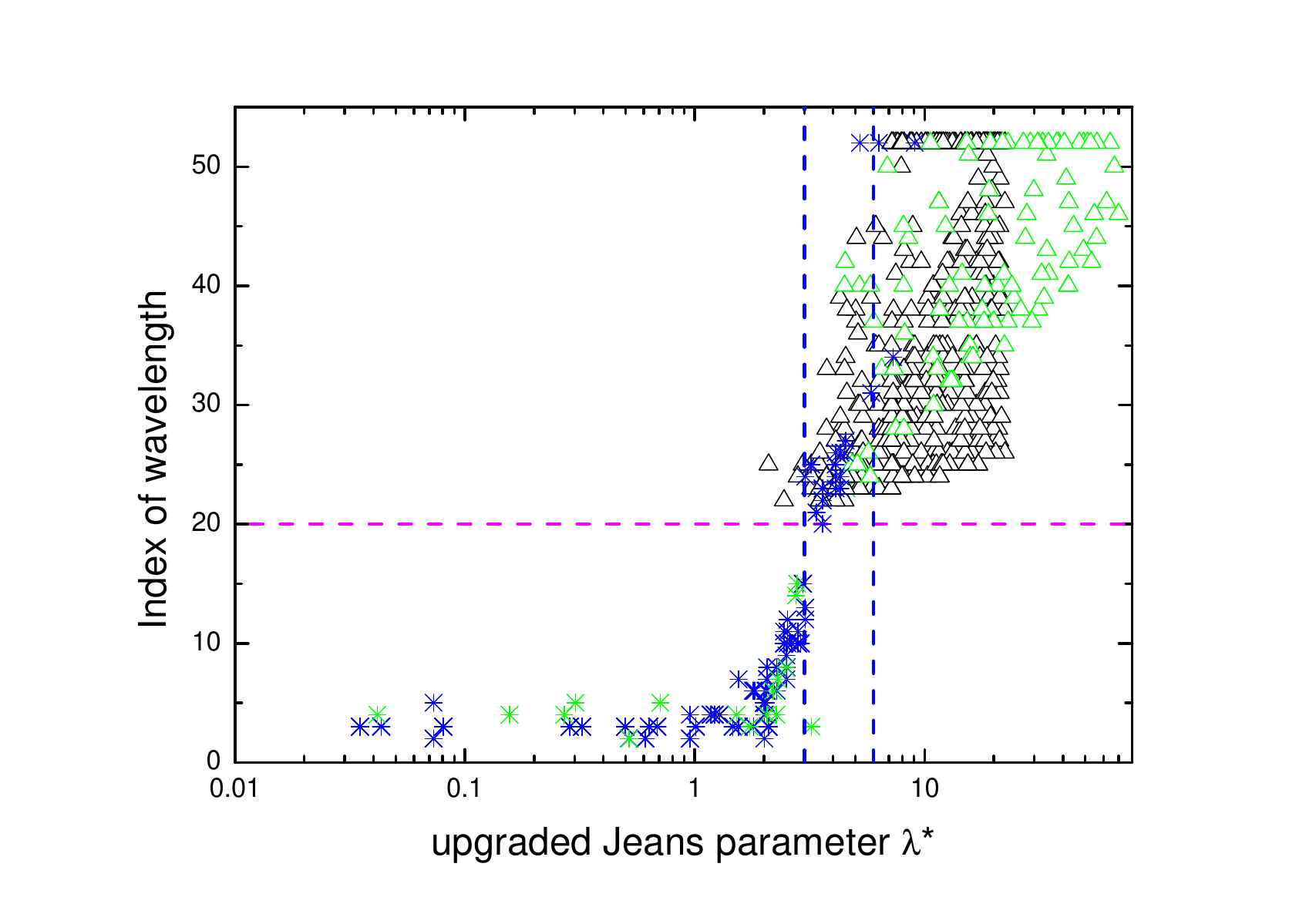}
\end{minipage}

\caption*{\textbf{Extended Data Fig. 4$\mid$The distribution of Rroche/Rs-Rxuv/Rs and the relationship between the location of sonic point and optical depthes of XUV spectrum.} Left panel: The dependence of the altitude distribution of the Roche radius, average absorption radius of XUV radiation and position of the sound point on the upgraded Jeans parameter. All planets orbiting G- and M-type are included.  ``red circle'': $\lambda^{*}<3$; ``green triangle'': $3<\lambda^{*}<6$; ``blue star'': $\lambda^{*}>6$. Right panel: \textbf{Photons that can reach the subsonic region.} The photon with different wavelength is characterized by an index. The number 1 represents the photon at wavelength of $1{\AA}$. The number 20 represents the wavelength of 100${\AA}$, and the number 53 represents a wavelength of 912${\AA}$. For X-ray, the wavelength is smaller than 100${\AA}$. Numbers higher than 20 indicate that the EUV photons can penetrate the subsonic region, and vice versa. The left ($\lambda^{*}=3$) and right ($\lambda^{*}=6$) vertical dashed lines separate the tidally-driven, tidally-XUV-driven transition, and XUV-driven regimes. The horizontal magenta dashed lines denote the index of wavelength at 100 ${\AA}$.}
\label{fidu_trans}
\end{figure}

\begin{figure}[tb]
\centering

\begin{subfigure}[t]{1.0\textwidth}
    \includegraphics[width=\linewidth,valign=t]{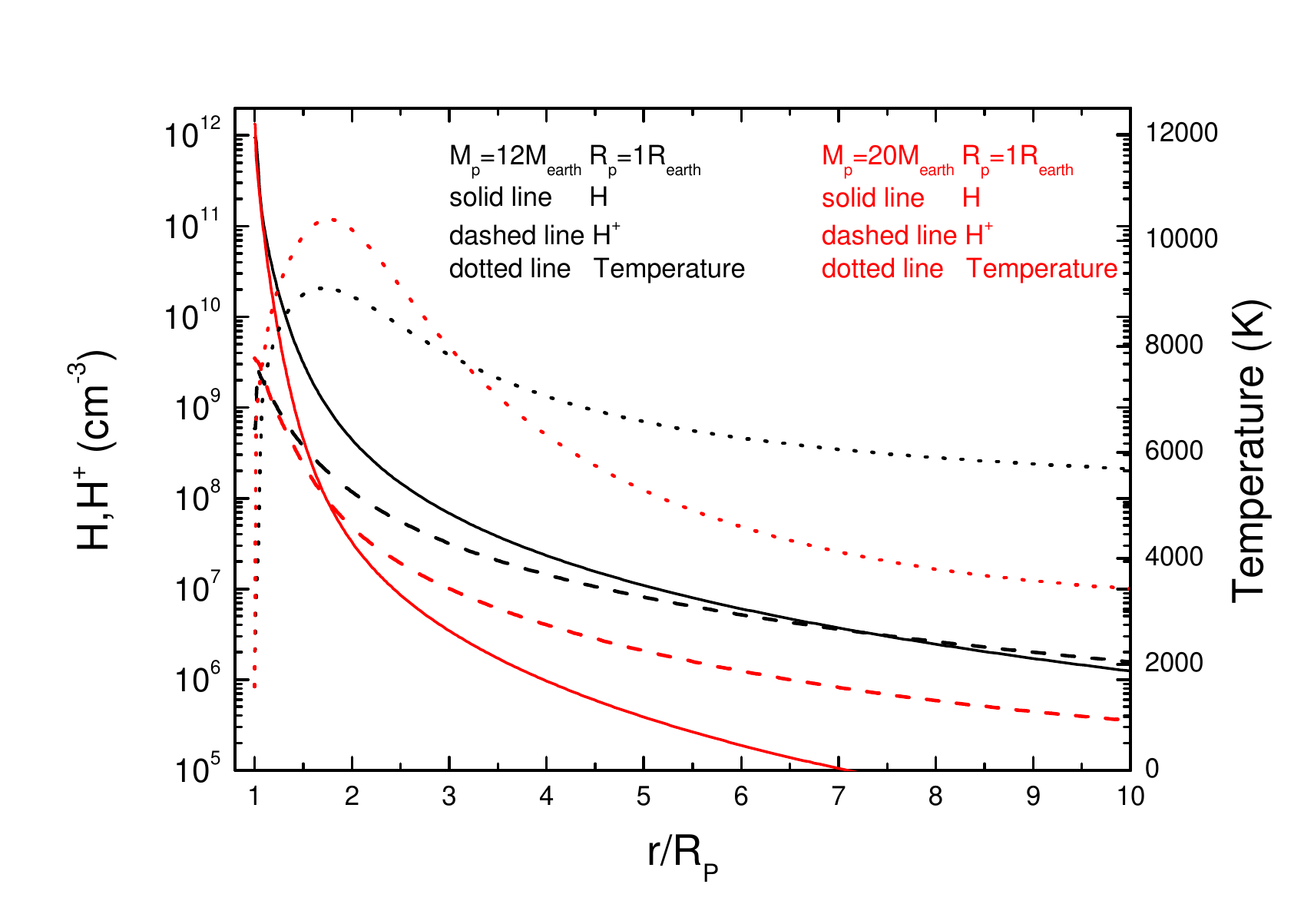}
%    \includegraphics{mass loss rate_F.pdf}
%    \label{Fig2a}
\end{subfigure}\hfill

\caption*{\textbf{Extended Data Fig. 5$\mid$ The structure of the atmosphere for two planets around the turning point of neutral and ionized wind.} The XUV flux received by them is 5.4$\times 10^{4} erg/s/cm^{2}$. The black lines denote the case of neutral atmosphere (M$_{p}=12M_{\oplus}$, R$_{p}=1R_{\oplus}$, and gravitational potential $\phi=7.50 \times 10^{12}$. ) while the red lines are the ionized atmosphere (M$_{p}=20M_{\oplus}$, R$_{p}=1R_{\oplus}$, gravitational potential $\phi=1.25\times 10^{13}$). `Solid line'': the number density of hydrogen atoms; `Dashed line'': the number density of hydrogen ions; `Dotted line'': the temperature profiles. }
\label{fidu_trans}
\end{figure}

\begin{figure}
\begin{minipage}[t]{0.5\linewidth}
\centering
\includegraphics[width=3.6in,height=2.5in]{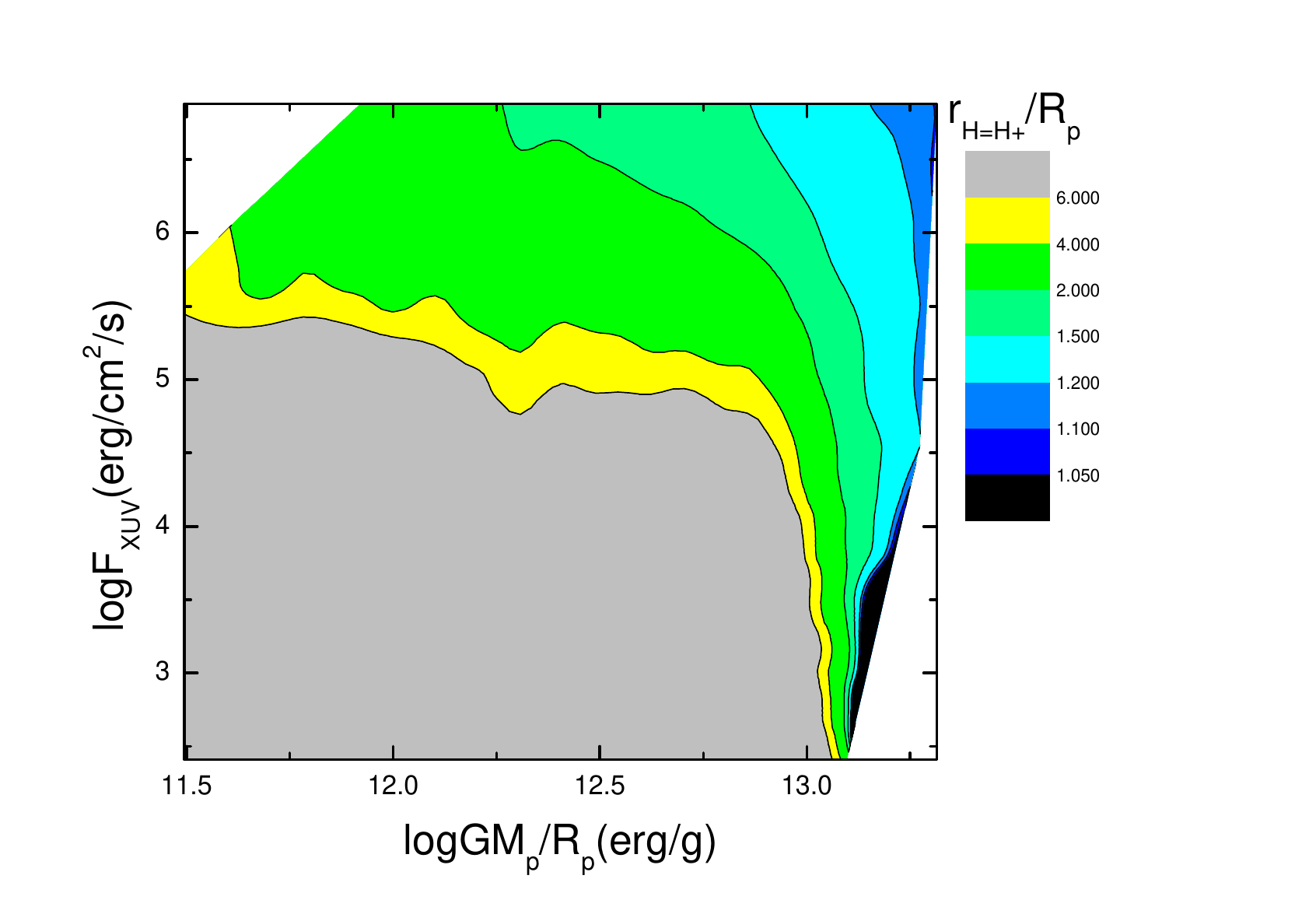}
\end{minipage}
\begin{minipage}[t]{0.5\linewidth}
\centering
\includegraphics[width=3.6in,height=2.5in]{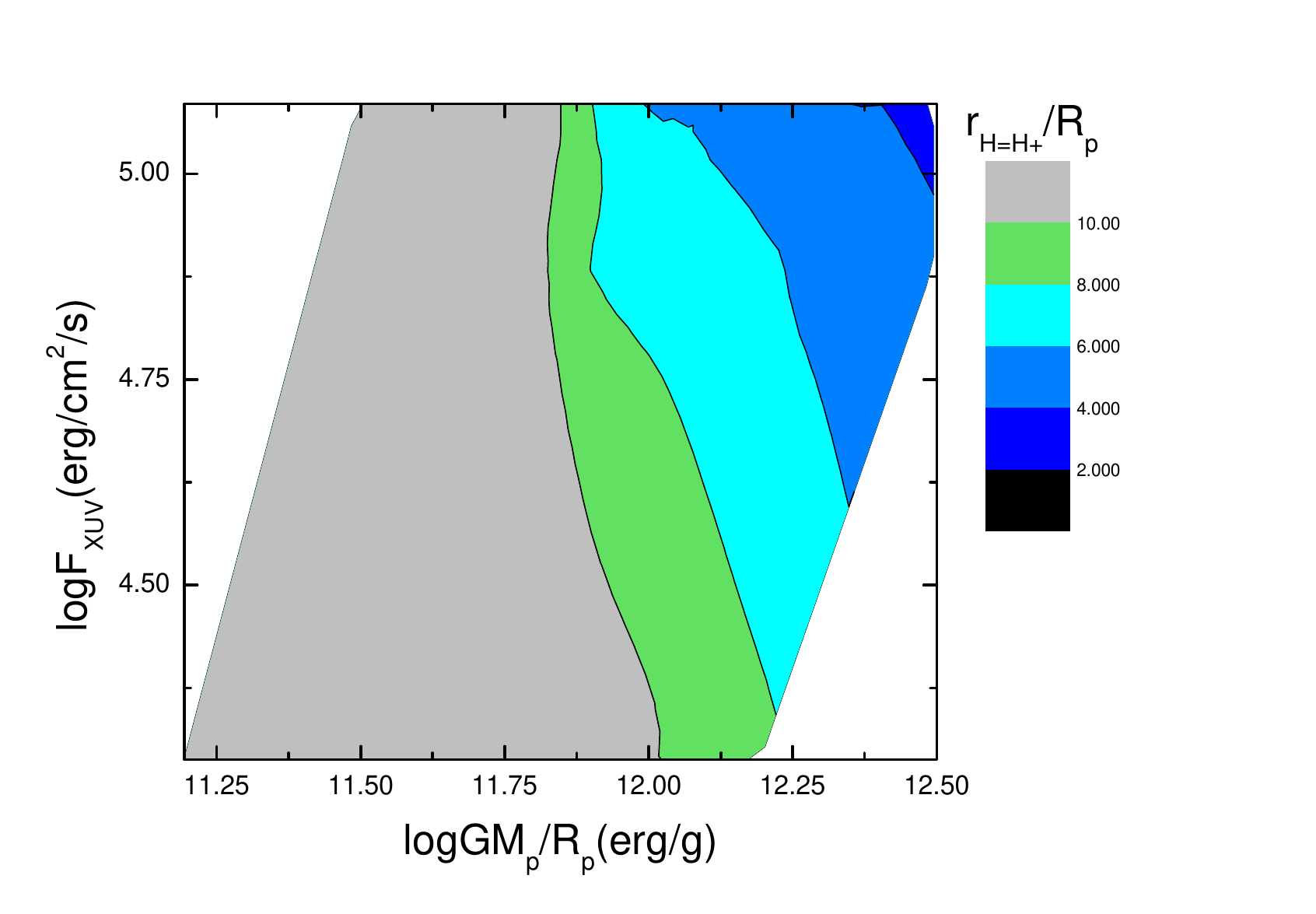}
\end{minipage}
\caption*{\textbf{Extended Data Fig. 6$\mid$The distributions of the radii of $n_{H}=n_{H}^{+}$.} Left panel: planets orbiting G-type star. Right panel: planets orbiting M-type star. At the locations, a transition of H/H$^{+}$ occurs.  The altitudes of R($n_{H}=n_{H}^{+}$) decrease with the decrease of F$_{XUV}$ and the gravitational potential. }
\label{fidu_trans}
\end{figure}

\begin{methods}

\subsection{The sample}
In this paper, I investigated the dependence of the different driving mechanisms on the physical parameters of the planet and star. Thus, a large sample of planets is calculated. The mass of the planet is set in the range of 1-30M$_{\oplus}$, which covers planets with the masses ranging from Earth-like planets to Neptune-like planets. Most of the planets of 1-30 earth masses have radii ranging from 1-4 R$_{\oplus}$, with a few exceptions having larger radii$^{50}$. For example, the masses of Kepler-51b and c are 2.1 and 4.0 (3.14) M${_\oplus}$ while their radii are 7.1 and 9.0 (8.98) R$_\oplus$, respectively $^{51-52}$. The reason for the large radius needs to be investigated further, but a H/He-rich envelope can result in an inflated radius $^{11,53}$. In this paper, the planetary radii of the calculation sample are set to increase from 1 to10 R$_{\oplus}$. In this situation, the densities of some planets are extremely low. However, the size of planets can be comparable with Bondi radius after the dissipation of protoplanetary disk $^{22}$. Therefore, the hydrodynamic escape behavior of these extremely low density planets is worth exploring. Our aim is to study the close-in exoplanets. Thus, the separations of my sample are less than 0.1au. Observations have revealed the correlation between the age and XUV radiation of star$^{54-55}$. In order to inspect the dependence of atmospheric escape on the XUV fluxes, I modeled the atmospheric escape at different ages. For the sake of completeness, I calculated two cases. First, the host star is sun-like star. The mass and the temperature of the host star are 1M$_{\odot}$ and 5760K, respectively. In fact, the temperature of the star changes with age. For G-type star, the variation is not dramatic in its main sequence phase. The parameters for planets orbiting a G-type star are concluded in Table 1. Note that not all samples can be calculated, specifically for the planets with very high gravitational potential and low XUV irradiation which together result in quasi-hydrostatic atmospheres. Finally, about 2050 planets are calculated successfully by using my models$^{31,41}$. Second, I also explored the escape of exoplanets orbiting M-stars because many exoplanets have been discovered around them. In this case, I calculated the escape of the atmosphere at separations of 0.02, 0.025 and 0.05AU. The mass and the temperature of the M-type stars are 0.5M$_{\odot}$ and 3000K, respectively. For M-type stars, I only calculated the case where the stellar age is 1Gyr. Thus, I did not consider the effect of evolution of the host star on the equilibrium temperature of planet. The planetary parameters are listed in Table 2.

\subsection{The model and boundary conditions}
I used the whole XUV spectra that is divided into 53 bins (1-912 ${\AA}$). Thus, my model is able to evaluate the heating by photons with different frequency correctly . I modified my 1D hydrodynamic model to calculate the structure of the escaping atmosphere. Specially, my early models$^{31,41}$ included the cooling of H$_{3}$$^{+}$, but neglected the cooling produced by Ly$\alpha$. The Ly$\alpha$ photons can diffuse out of the wind in a typical mass loss rate of 10$^{10}$ g/s$^{5}$. For the sake of completeness, here I included the Ly$\alpha$ cooling as was done by other authors$^{5,23,30}$. In addition, I also included the cooling produced by metal atoms O, C and Si (for details see Table. 5 of reference 56). The magnitude of pressure on the surface of a planet depends on the composition of the atmosphere. For a planet with a clear atmosphere, the pressure on its surface of the planet can range from several to one hundred mbar$^{57}$. However, the photochemical hazed model suggest that a young and low mass planet (super-puffs) has a hazed atmosphere. The pressure at the planetary surface is only 10-100 nbar$^{58}$. In my calculation, the temperature and the pressure at planetary radius are set as the equilibrium temperature and 1 $\mu$bar. In fact, the number density ($\sim$4$\times 10^{12}- 10^{13}/cm^{3}$) corresponding to 1 $\mu$bar is sufficient to absorb the XUV radiation$^{5,24,48}$. Therefore, the modelling results are not affected by the value of the lower boundary pressure I set. I further tested a situation where the lower boundary pressure is 100 mbar, and found that the TE-driven escape still occurs if $\lambda<\sim3$, otherwise there are no the TE-driven escape. Therefore, the magnitude of pressure of the lower boundary does not affect the conclusion about the classification of driving mechanism. I set the solar chemical composition, and H$_{2}$ is dominant (the number densities of H and H$^{+}$ are much smaller than that of H$_{2}$) in the lower boundary because they are lower than 2000K (The H$_{2}$ can be decomposed if the temperature is higher than 2000K). I found that the case with $\lambda<3$ are unstable when the bulk velocity at the lower boundary is higher than the sound speed. Therefore, I imposed the velocity of the lower boundary to equal with the sound speed while it is larger than the sound speed. This confines the mass loss rate because in the condition, the velocity of the atmosphere at the lower boundary is given. In order to obtain a stable solution of the TE-driven escape for those extremely inflated planets, sometimes the pressure at lower boundary needs to be decrease to lower values ($\sim$ 0.1 $\mu$bar). In this condition, I found that the mass loss rates are in the orders of magnitude of $\sim$ 10$^{14}$-10$^{15}$ g/s, which is far greater than those of XUV driven escape. Therefore, the boundary condition imposed has no impact on the result of classification (See the section of The TE-driven escape in the Method. ).

%for planets with a compacted atmosphere (the cases of high $\lambda$), the lower boundary is slightly higher than the planetary radius owing to the %small scale height even if the atmospheres have no hazes so that R$_{p}\sim $R$_{lowerboundary}$.
\subsection{The evolution of XUV emission}
The XUV radiation is related to the activity of the star, which is modulated further by the rotation. Thus, the XUV radiation will decrease with age because the spin down of the magnetic braking of stellar wind. Here I used the empirical relation of Sanz-Forcada et al.$^{54}$ to produce the X-ray and EUV luminosity of host stars. The evolution of the luminosity of X-ray and EUV with age can be expressed as:

\begin{equation}
L_{x}=\left\{
     \begin{array}{lcl}
      6.3\times10^{-4}L_{bol} (t \leq t_i)  \\
      1.89\times10^{28}t^{-1.55}(t \textgreater t _i)
      \end{array}
      \right.
\end{equation}

\begin{equation}
log L_{euv}=(29.12\pm 0.11)-(1.24\pm 0.15)log\emph{t}
\end{equation}
with $\emph{t}_{i}$=2.03$\times10^{20}L_{bol}^{-0.65}$, where $\emph{t}$ is the age of the star in Gyr. L$_{X}$ and L$_{euv}$ are the luminosity (erg s$^{-1}$) in the X-ray and EUV band, respectively. In this paper, I investigated the hydrodynamic escape with age of 0.05, 0.5, 1.0, 2.4, 4.6 and 8Gyr. With the decrease of the age, the L$_{X}$/L$_{XUV}$ varies from $\sim$23\% to $\sim$11\%. Thus, one can expect that the effect of X-ray will become unimportant for planets around old stars. I used the software XSPEC-APEC to obtain the XUV spectra. The free parameters include the stellar atmospheric metal abundance and the temperature in the coronal layer. Here, I first set the metal abundance to be the solar abundance so that the output SED is only related to the temperature. Furthermore, I calculated lots of spectra and selected those theoretical spectra as the input spectra by comparing the theoretical and empirical ratios of L$_{X}$ and L$_{euv}$ $^{54}$.

\subsection{Role of X-ray}
I have used the fitting results of Ricotti$^{59}$ and Dalgarno$^{60}$ et al. to calculate the net heating efficiency and secondary ionization of H, H$_{2}$ and He in the range of 50-1000${\textmd{\AA}}$ (for detail see reference 60). The net heating efficiency is a function of both wavelength and altitude because they are related to the electron abundance fraction in the gas and the wavelength of incident photons. In order to inspect the influence of X-ray on the mass loss, in the paper the wavelength range is extended to 1-912${\textmd{\AA}}$. Cecchi-Pestellini et al.$^{61}$ calculated the heating efficiencies for X-ray photons with energy 300ev (40${\textmd{\AA}}$) and 1Kev (1.2${\textmd{\AA}}$) at different values of the electron mixing ratio. Thus, for the range of 1-40${\textmd{\AA}}$, I fitted their heating efficiencies of different electron mixing ratio (x$_{e}$) and interpolated the heating efficiencies at different altitudes. I used the results of x$_{e}$=0.1 when the electron mixing ratio is larger than 0.1 because the maximum electron fraction of Cecchi-Pestellini et al. is 0.1. In fact, owing to their small cross sections only the high energy photons penetrate to the bottom of the atmosphere where the ionization degree is low in general. Thus, using the results of x$_{e}$=0.1 is reasonable generally.

Owen \& Jackson $^{12}$ used single photon frequency to represent the irradiation in both X-ray and EUV bands. This approach overlooked the fact that photons of different frequencies have varying penetration depths. Among the planets that experience XUV-driven atmospheric escape, I conducted a further investigation to determine if X-ray alone is capable of driving an escape as predicted by Owen \& Jackson$^{12}$. To do this, I compared the R$_{\lambda}(\tau_{\lambda}=1)$ and R$_S$. The calculation results showed that for all XUV-driven escapes, EUV photons can penetrate the subsonic region (right panel of Extended data Fig. 4). Therefore, the X-rays cannot solely drive an escape. An intriguing phenomenon is that for most planets with tidal-driven escape ($\lambda^{*}<$3), only X-ray can penetrate the subsonic region, while for the remaining planets ($3<\lambda^{*}<6$), EUV can also penetrate the subsonic region. For these planets, however, atmospheric escape is dominated by tidal forces.

\subsection{The TE-driven escape}

In the section, I verified that the escape of the atmosphere is driven solely by the thermal energy of planet itself if the value of $\lambda$ at the lower boundary is smaller than $\sim$3. As expressed by many works $^{5,62}$, the hydrodynamic equations can be converted to an alternative form for the velocity via some manipulations. Here, I used the equation (17) of Garcia Munoz$^{62}$ to express this, namely
\begin{equation}
\frac{\partial \ln u}{\partial \ln r}=\frac{-2-\frac{\lambda}{\gamma}\frac{f_{ext}}{G M_{p}/r^{2}} +(\gamma-1)\frac{r}{c^{2}\textrm{F}}(r^{2}Q-\frac{\partial}{\partial r}(r^{2}q))}{1-M^{2}}
\end{equation}
where M=u/u$_{sound}$ is the Mach number, u$_{sound}$ is the sound speed, $f_{ext}$ is the external forces, Q is the heating term, q is the heat flux, and F=$\rho u r^{2}$. For the TE-driven escape, the heating term and the external forces other than the planetary gravitational force can be ignored. Furthermore, the heat flux can also be omitted. In such situation, the equation can be simplified to:
\begin{equation}
\frac{\partial \ln u}{\partial \ln r}=\frac{-2+\frac{\lambda}{\gamma}}{1-M^{2}}
\end{equation}

The equation above means that the signs of the numerator and denominator should be same if an atmosphere expands outward monotonously. Otherwise, the atmosphere will be decelerated. The critical point is defined when $\lambda_{c}=2\gamma$. The $\gamma$ is 5/3 or 7/5 for mon- and diatomic gases. Thus, this requires $\lambda_{c} \simeq 3.3$ or 2.8. In addition, the TE-driven escape requires that the sum of kinetic energy and enthalpy per unit mass is greater than the gravitational potential energy per unit mass$^{63}$, namely, $\frac{u_{0}^{2}}{2}+\frac{\gamma}{\gamma-1}\frac{k_{B}T}{\mu}>\frac{GM_{p}}{R_{p}}$, where $u_{0}$ is the velocity of the lower boundary. In this situation, the Jeans parameter at the lower boundary has $\lambda_{b}<\frac{u_{0}^{2}}{2a_{0}^{2}}+\frac{\gamma}{\gamma-1}$, where $a_{0}^{2}=\frac{k_{B}T}{\mu}$.

There are two cases at the lower boundary for the solution of Equation (6), one of which is subsonic and the other is supersonic (or equal to the sound speed). For the former, the corresponding solution in the entire atmosphere is subsonic or transonic, while for the latter, it is supersonic. If the Jeans parameter at the lower boundary ($\lambda_{b}$) is smaller than $\lambda_{c}$, a transonic solution requires the value of $\lambda$ at the sonic point is equal to $\lambda_{c}$ and is larger than $\lambda_{b}$, which further requires a declining temperature profile to counteract the decrease of $\lambda$ due to the increase of radius in the decelerating outflow ($\frac{\partial \ln u}{\partial \ln r}$ is negative in the subsonic regime for an arbitrary subsonic velocity at the lower boundary). Here, the TE-driven escape is adiabatic. For a decelerating outflow, the work done by PdV (adiabatic compression) is used to increase the internal energy. Thus, the temperatures of the atmosphere in the subsonic regime will continue to rise. As a result of the increase of the temperature, the $\lambda$ is always smaller than $\lambda_{c}$ with the increase of radius, which conflicts with the condition of transonic solution as mentioned above. Thus, the transonic solution is impossible. The subsonic solution in the entire atmosphere is also impossible because of the same reason. Thus, only the supersonic solution is appropriate if $\lambda_{b} < \lambda_{c}$, which hints that the velocity at the lower boundary should be greater than the sound speed. The supersonic TE-driven escape requires $\lambda_{b} <$ min$(2\gamma$, $\frac {u_{0}^{2}}{2a_{0}^{2}}+\frac{\gamma}{\gamma-1})$ and $u_{0}^{2} \geq \gamma a_{0}^{2}$. Due to $\frac {u_{0}^{2}}{2a_{0}^{2}}+\frac{\gamma}{\gamma-1}\geq 2\gamma$, the supersonic escape is further constrained by the condition of $\lambda_{b} <2\gamma$, namely, $\lambda_{b} <3.3$ or $<$2.8 for mon- and diatomic gases, respectively.

If $\lambda_{b} > \lambda_{c}$, it need to fulfill the condition of $\frac{u_{0}^{2}}{2a_{0}^{2}}+\frac{\gamma}{\gamma-1}>\lambda_{b} >2\gamma $ which hints that a large $\lambda_{b}$ also seems to meet the conditions for supersonic escape throughout the atmosphere as long as u$_{0}^{2}$ is much larger than $\gamma$a$_{0}^{2}$. However, the condition of u$_{0}^{2} >\gamma$ a$_{0}^{2}$ for planets with high $\lambda_{b}$ is not predicted by previous works$^{30,57}$ so that the supersonic solution can be excluded. Thus, there can exist subsonic or transonic solutions if $\lambda_{b} > \lambda_{c}$. The condition means $3.3>\lambda_{b}\geq3.3$ and $4.2>\lambda_{b}\geq2.8$ for mon- and diatomic gases in the limit of $u_{0}=\sqrt{\gamma}a_{0}$, respectively. It is clear that there is no subsonic or transonic escape for monatomic gases.
In addition, $u_{0}$ will be much smaller than $a_{0}$ when $\lambda_{b}$ approaches $\frac{\gamma}{\gamma-1}$ (see below). Therefore, the condition of solution can be expressed as $\frac{\gamma}{\gamma-1}>\lambda_{b} >2\gamma $. For diatomic gases, the condition is $3.5>\lambda_{b} >2.8$.

Such behavior has also been verified partially by previous hydrodynamic calculation$^{48}$. I also justify this behavior in my calculations by ignoring the tidal forces, XUV irradiation and all chemical reactions. As shown in Extended data Fig. 2, there is a boundary around $\lambda\approx$3. The mass loss rates of the planets in the regions of $\lambda<$2.8 are in the order of magnitude of 10$^{15}$-10$^{17}$ g/s and their velocities at the lower boundary are supersonic (see Extended data Fig. 3a. Note also that the velocities at the lower boundary are not imposed to be equal with the sound speed). In the regime of $2.8<\lambda<3.5$, the escape rates of some planets remain high, though their velocities at the lower boundary turn to be subsonic. The closer the value of $\lambda$ approaches 3.5, the lower the velocity at the lower boundary and the corresponding mass loss rate. Theoretically, the minimum mass loss rate for transonic solutions occurs when the sonic point coincides with the exobase . However, our numerical model encountered numerical instability when $\lambda$ extreme approached 3.5 so that the lowest value of the mass loss rate ($\sim10^{11}$g/s) predicted here can be higher than that of the most extreme case. In the range of $\lambda_{b}<3.5$, the mass loss rates decline with the increase of $\lambda$. If $\lambda_{b}>3.5$, there is no hydrodynamic solution when the tidal forces and XUV irradiation are neglected. The results of numerical model are completely consistent with the boundary of atmospheric escape predicted above. Furthermore, the mass loss rates of specific planets with $\lambda_{b}<3$ are independent of the tidal forces and the XUV radiation of the host star if the lower boundary velocity is forced to be equal with the speed of sound. For example, the mass loss rates of the planet with 1 M$_{\oplus}$ and 4R$_{\oplus}$ at 0.03au ($\lambda_{b}$ and $\lambda_{b}^{*}$ are 2.72 and 0.65, respectively) are 3.08$\times$10$^{14}$ g/s whatever the F$_{XUV}$ is. Furthermore, the mass loss rate for the case of no XUV radiation and tidal forces is 3.01$\times$10$^{14}$ g/s that is identical with the former although the distribution of the densities, velocities and the temperatures shows significant differences (see the panels of Extended data Fig. 3b and 3c. The slight difference in mass loss rates between the two is attributed to the small pressure difference at the lower boundary).

Fossati et al.$^{57}$ suggest that for $\lambda <$ $\sim$ 15-35, hydrodynamic escape is driven by internal thermal energy, namely, the XUV radiation is unimportant in this regime. However, the total energy (the sum of the kinetic energy and the enthalpy minus the potential energy) per unit mass at the lower boundary is negative if $3.5<\lambda <\sim 15-35$. Therefore, an escape requires external energy sources, such as XUV heating, to provide enough energy. In the situation, the escape for a planet is triggered by external energy sources though the thermal energy also play a role, which is different from the TE-driven escape defined in this paper. For the latter, the escape is driven solely by planet's own thermal energy, namely, XUV heating or the work done by the stellar tidal forces is not necessary.

\subsection{Distributions of R$_{XUV}$, R$_{Roche}$ and R$_{s}$}

The mean absorption radius of XUV photons is defined as: $R_{XUV}=\frac{\sum_{\lambda}R_{\lambda}(\tau_{\lambda}=1)F_{\lambda}}{F_{XUV}}$, where R$_{\lambda}(\tau_{\lambda}=1)$ is the radius at which the optical depth is unity for a given wavelength $\lambda$. F$_{\lambda}$ and F$_{XUV}$ are the XUV flux at wavelength $\lambda$ and the integrated flux across the entire XUV range, respectively. I plotted all groups in the  R$_{XUV}$/R$_{s}$-R$_{Roche}$/R$_{s}$ plane in left panel of Extended data Fig. 4. One can expect that the atmospheric escape of the planets above the line of R$_{Roche}$=R$_{XUV}$ is mainly driven by tidal forces because the altitudes of R$_{XUV}$ are higher than R$_{Roche}$. Our calculation results also show that in the regions above the line of R$_{Roche}$=R$_{XUV}$, the corresponding $\lambda^{*}$ values of most planets are smaller than 3 although some planets have values between 3 and 6. For those planets below the line of R$_{Roche}$=R$_{XUV}$, the R$_{XUV}$ is closer to the surface of planet than R$_{Roche}$. The atmospheric escape is dominant by XUV heating. Specifically, the planets distribute in four quadrants. The XUV absorption radii of those planets in the first quadrants are distributed in a wide range and are greater than their Roche radii (R$_{Roche}<$R$_{XUV}$). Therefore, their escapes are driven by tidal forces and the values of $\lambda^{*}$ of most planets are smaller than 3. At the same time, there are almost no planets if R$_{s}<$R$_{XUV}<$R$_{Roche}$. My results show that neither the heating of XUV nor the tidal forces of the star can drive an escape if R$_{XUV}$ is smaller than R$_{Roche}$ but is much larger than R$_{s}$. The fourth quadrant is where the XUV absorption radii are lower than the sonic point radii, but the Roche radii are higher than the altitudes of the sonic point. As expected, most values of $\lambda^{*}$ in the quadrant exceed 6, and the hydrodynamic escape of planets with $\lambda^{*}>$ 6 is actually controlled by XUV heating. In the second and third quadrants, most planets are distributed around the line of R$_{Roche}$=R$_{XUV}$. In fact, the closer a planet approaches the line of R$_{Roche}$=R$_{XUV}$, the harder it becomes to distinguish its escape mechanism. In this situation, the driving mechanism around the regions where R$_{Roche}\approx$R$_{XUV}\approx$R$_S$ is ambiguous and could be either tidally- or XUV-driven. Our results also indicate that the values of $\lambda^{*}$ around the regions of R$_{Roche}$=R$_{XUV}$ are distributed in the range of 3-6. In addition, some planets in the second and third quadrants are far from the line of R$_{Roche}$=R$_{XUV}$. Their corresponding values of $\lambda^{*}$ are smaller than 3 or greater than $\sim$ 6, respectively. Therefore, the driving mechanisms can be well distinguished in the regimes of $\lambda^{*}$.

\subsection{The energy-, photo- and recombination-limited regimes}

In XUV-driven regime, Murray-Clay et al.$^{5}$ found that the mass loss rate is limited by the XUV energy deposited in the atmosphere or the radiation-recombination equilibrium of the escaping gas when the XUV irradiation changes from low to high. The two types of mass loss are termed energy- and recombination-limited escape. In addition, Owen \& Alvarez$^{42}$ found that under the moderate irradiation, the mass loss rates of low gravity planets are determined by the number of incoming ionising photons. In this situation, the mass loss rate scales to the square of the effective absorption radius of the planet. It is photo-limited escape.

Lamp$\acute{o}$n et al. $^{64}$ reported that the regions of ionization front (IF) of recombination- energy- and photon-limited escape change from narrow to wide for hot Neptunes and Jupiters. The ionization front stands for the region where $0.05 \leq f_{H^{0}} \leq 0.95$ (f$_{H^{0}}$ is the mole fractions of atomic hydrogen). Here, the position of $n_{H}/n_{H^{+}}$=1 can roughly represent the central point of IF. According to Lamp$\acute{o}$n et al, the IF of recombination-limited escape is very narrow, which means that the altitude of $n_{H}/n_{H^{+}}$=1 is close to the planetary radius. The altitudes of $n_{H}/n_{H^{+}}$=1 should be increasing from the energy- to photon-limited escape. From my calculation results, the locations of $n_{H}/n_{H^{+}}$=1 of the ionized winds are generally lower than those of neutral winds. For instance, I checked the profiles of H, H$^{+}$ and temperature of specific planets with the same F$_{XUV}$ on the both sides of the ionized-neutral boundary. As expected, the planets with higher gravitational potential show lower altitudes of $n_{H}=n_{H^{+}}$ than those of lower gravitational potential (see Extended data Fig. 5). The locations of $n_{H}=n_{H^{+}}$ are about 7.5R$_{p}$ and 1.7R$_{p}$ for the cases of low and high gravitational potentials, which means that from the left to right of the neutral-ionized boundary, a transition among the photo-, energy- and recombination-limited escape can occur. Moreover, by comparing with Fig. 1 of Owen \& Alvarez$^{42}$, the planets with the ionized wind partially overlap with the energy- and recombination-limited regime. However, the overlap is dependent of the XUV fluxes and the heating efficiencies assumed in their model.

To further describe these three regimes, I plotted the altitudes of $n_{H}=n_{H^{+}}$ of all XUV-driven planets in Extended data Fig. 6. According to Lamp$\acute{o}$n et al, the escape is roughly classified as energy-limited if $1.1<\frac{r_{H=H^{+}}}{R_{p}}<\sim 1.5$. It is recombination- or photo-limited escape if $\frac{r_{H=H^{+}}}{R_{p}}<\sim 1.1$ or $\frac{r_{H=H^{+}}}{R_{p}}>\sim 1.5$. I emphasized that an accurate determination for these three regimes requires a comparing between recombination and hydrodynamic timescales $^{64}$, which is beyond the scope of this work. For planets around G-type stars, the planets with high gravitational potential can locate at the recombination-limited regime (see left panel of Extended data Fig. 6). With the decrease of gravitational potentials and the XUV fluxes, the locations of $n_{H}=n_{H^{+}}$ increase. The gravitational potential distribution of energy-limited ranges from medium to high, and the XUV radiation from stars to them is also distributed in a wide range (blue and dark green areas of left panel). For planets orbiting a M-type star, the energy- and recombination-limited escapes do not occur (Note that the XUV radiation of the M-type star is in a middle level because I only calculated the case that the age of the star is 1 Gyr. For planets orbiting younger M-type stars, these two types may occur). Finally, the planets with lower gravitational potential and XUV radiation are in the regime of photon-limited escape (roughly, the location of $n_{H}/n_{H^{+}}$ is greater than 2 times planetary radius), which means that most of the planets around G-type stars and all of the planets orbiting M-type stars are this regime.

\subsection{The criterion of supersonic and subsonic escape}

Johnson et al.$^{46}$ used the sonic boundary to define the minimum value of the net heating rate (Q$_{net}$) in hydrodynamic model, namely, the energy deposited in the atmosphere can not driven a transonic outflow if the net energy absorbed from the XUV radiation is smaller than the critical value Q$_{c}=4\pi r_{s}\frac{\gamma}{\sigma_{c}K_{n}}\sqrt{\frac{2U(r_{s})}{\mu}}U(r_{0})$, where subscript $\emph{s}$ denote the sonic point. U$(r_{s})$ and U(r$_{0}$) are the gravitational energy at the sonic point and the lower boundary, respectively. $\sigma_{c}$ is the collision cross section. In neutral atmosphere, the collision cross section of H-H$^{+}$ ($\sigma_{H-H^{+}}$=2 $\times$10$^{-15}$ cm$^{2}$)$^{65}$ is two orders of magnitude larger than that of H-H ($\sigma_{H-H}$=10$^{-17}$ cm$^{2}$)$^{65}$. In ionized atmosphere, the process of H$^{+}$-H$^{+}$ ($\sigma_{H^{+}-H^{+}}$= $10^{-13}(T/10^{4} K)^{-2}$cm$^{2}$)$^{5}$ dominates the collisions. Thus, Q$_{c}$ defines a minimum net heating rate that can drive a transonic outflow in the collision-dominated regime. Johnson et al.$^{46}$ define that the sonic point will occur in the collision-dominated regime if Q$_{net}>$ Q$_{c}$. Note that the hydrodynamic model is not applicable above the exobase. To connect the collision and collisionless atmosphere, a better method is to find the exobase firstly, and the modified Jeans condition is applied in the upper boundary if the exobase occurs in the subsonic regions$^{3,66}$. However, I emphasize that applying hydrodynamic model throughout whole regions will not affect the classification about supersonic or subsonic escape.

\end{methods}

%% Put the bibliography here, most people will use BiBTeX in
%% which case the environment below should be replaced with
%% the \bibliography{} command.

\begin{addendum}
\item I thank the anonymous reviewers for their constructive comments and language polishing on this paper, which helped improve the manuscript. This work is Supported by the Strategic Priority Research Program of Chinese Academy of Sciences,Grant No. XDB 41000000, the National Key R\&D Program of China (grant No. 2021YFA1600400/2021YFA1600402), and National Natural Science Foundation of China (Nos. 12288102 and 11973082). The authors gratefully acknowledge the “PHOENIX Supercomputing Platform” jointly operated by the Binary Population Synthesis Group and the Stellar Astrophysics Group at Yunnan Observatories, Chinese Academy of Sciences.

\end{addendum}

 \begin{table}
 \caption{The planet parameter orbiting solar-like stars in the grid calculations.}
% \label{tab:par}
% \begin{tabular}{@{}ccccccccc}
 \begin{tabular}{cccccccccccc}
% \begin{tabular}{cccc}
  \hline
Mass (M$_{\oplus}$) &1 &2& 4 & 8& 12& 16& 20& 25&30\\
\hline
Radius(R$_{\oplus}$)&1 & 1.5& 2& 3& 4& 6 & 8& 10& &\\
\hline
Separation (AU) & 0.02& 0.03& 0.05&0.075&0.1 \\
Teq (K)       &1935& 1580& 1224& 999&866\\
\hline
age (Gyr) & 0.05  & 0.5 & 1  & 2.4 &4.6 &8.6 & &  \\
L$_{XUV}$($\times 10^{27}$erg/s)  &8560&367&137&39.9&15.9&7.29& & &\\
  \hline
 \end{tabular}
\end{table}

 \begin{table}
 \caption{The planet parameter orbiting M-type stars in the grid calculations.}
% \label{tab:par}
% \begin{tabular}{@{}ccccccccc}
 \begin{tabular}{cccccccccccc}
% \begin{tabular}{cccc}
  \hline
Mass (M$_{\oplus}$) &1 &2& 4 & 8& 12& 16& 20& 25&30\\
\hline
Radius(R$_{\oplus}$)&1 & 1.5& 2& 3& 4& 6 & 8& 10& &\\
\hline
Separation (AU) & 0.02& 0.025& 0.05& & \\
Teq (K)       &850& 760& 540&  &\\
\hline
age (Gyr) & 1  &  &   & & && &  \\
L$_{XUV}$($\times 10^{27}$erg/s)  &137 & && & & & & &\\
  \hline
 \end{tabular}
\end{table}

%%
%% TABLES
%%
%% If there are any tables, put them here.
%%

\end{document}